\pdfoutput=1
\documentclass[sigplan,10pt,nonacm]{acmart}

\acmConference[PPoPP '27]{The 32nd ACM SIGPLAN Annual Symposium on
  Principles and Practice of Parallel Programming}{March 20--24, 2027}{Salt
  Lake City, UT, USA}
\acmYear{2027}
\copyrightyear{2027}
\setcopyright{none}          %
\acmDOI{}
\acmISBN{}

\usepackage{subcaption}       %
\usepackage{listings}         %
\usepackage{tikz}             %
\usepackage{pgfplots}         %
\pgfplotsset{compat=1.18}
\usetikzlibrary{positioning,arrows.meta,fit,backgrounds,calc}

\definecolor{fsblue}{RGB}{47,84,150}
\definecolor{fsorange}{RGB}{214,116,28}
\definecolor{fscell}{RGB}{136,166,208}

\graphicspath{{figures/}}

\usepackage{xspace}   %

\newcommand{\sysname}{ForgeStencil}
\newcommand{\kagent}{Kernel~Agent}
\newcommand{\aagent}{App~Agent}
\newcommand{\harness}{\sysname{} Harness}

\newcommand{\forge}{\emph{Forge Engineering}\xspace}

\newcommand{\numapps}{100}              %
\newcommand{\numappstotal}{116}         %
\newcommand{\medspeedupnum}{1.41}       %
\newcommand{\kernelgmnum}{2.35}         %
\newcommand{\kernelfphgmnum}{1.95}      %
\newcommand{\medspeedup}{\medspeedupnum$\times$}  %
\newcommand{\gmspeedup}{2.05$\times$}   %
\newcommand{\kernelgm}{\kernelgmnum$\times$}    %
\newcommand{\kernelmed}{2.83$\times$}   %
\newcommand{\kernelfphgm}{\kernelfphgmnum$\times$} %
\newcommand{\kernelfphmed}{1.67$\times$}%
\newcommand{\kernelallgm}{2.08$\times$} %
\newcommand{\varcoeffgm}{1.34$\times$}  %
\newcommand{\rooflinepeak}{74--85\%}    %
\newcommand{\rooflinemedian}{76.6\%}    %
\newcommand{\numkernels}{81}            %
\newcommand{\numdispatchconfigs}{41}    %
\newcommand{\numdispatchkernels}{15}    %
\newcommand{\numrejected}{16}           %
\newcommand{\numretracted}{5}           %
\newcommand{\numfused}{13}              %

\newcommand{\tcutil}{11\%}              %
\newcommand{\tcconvloss}{1.39$\times$} %
\newcommand{\boxdensitypenalty}{6.4}   %

\newcommand{\kernelci}{[1.86, 2.92]}      %
\newcommand{\arbwall}{0.87$\times$}       %
\newcommand{\arbncu}{1.63$\times$}        %

\newcommand{\genshaperatio}{0.98}         %
\newcommand{\gencubicgm}{4.11$\times$}    %
\newcommand{\genedgegm}{4.04$\times$}     %
\newcommand{\genappmed}{1.33$\times$}     %
\newcommand{\genappgm}{1.59$\times$}      %

\newcommand{\ablforgelo}{2.29$\times$}    %
\newcommand{\ablforgehi}{2.73$\times$}    %
\newcommand{\ablautolo}{1.06$\times$}     %
\newcommand{\ablautohi}{1.28$\times$}     %
\newcommand{\abloneshotlo}{0.85$\times$}  %
\newcommand{\abloneshothi}{0.87$\times$}  %

\newcommand{\archbasegm}{1.81$\times$}      %
\newcommand{\archbaselo}{1.57$\times$}      %
\newcommand{\archbasehi}{2.11$\times$}      %
\newcommand{\archforgeboxa}{2.29$\times$}   %
\newcommand{\archforgeboxh}{2.06$\times$}   %
\newcommand{\archforgediaa}{2.73$\times$}   %
\newcommand{\archforgediah}{2.15$\times$}   %
\newcommand{\archhalidea}{2.16$\times$}     %
\newcommand{\archhalideh}{2.38$\times$}     %
\newcommand{\archebisulo}{1.03$\times$}     %
\newcommand{\archebisuhi}{1.08$\times$}     %

\newcommand{\archsotagmh}{2.16$\times$}     %
\newcommand{\archsotacells}{17}             %
\newcommand{\bgencells}{41}                 %
\newcommand{\bgengm}{1.51$\times$}          %
\newcommand{\bgenbw}{2.39$\times$}          %
\newcommand{\bgenforgehi}{7.0\%}            %
\newcommand{\bgenrounds}{18}                %
\newcommand{\bgenocch}{5.5\%}               %
\newcommand{\bgenocca}{0.6\%}               %

\newcommand{\eg}{e.g.,\xspace}

\newcommand{\etal}{et~al.\xspace}

\begin{document}

\title{ForgeStencil: Automating Per-Case Stencil Specialization from
  Kernels to 100+ Real Applications}

\author{Yaojian Chen}
\affiliation{%
  \institution{Tsinghua University}
  \city{Beijing}
  \country{China}
}

\author{Yuxuan Li}
\affiliation{%
  \institution{Tsinghua University}
  \city{Beijing}
  \country{China}
}

\author{Wubing Wan}
\affiliation{%
  \institution{Tsinghua University}
  \city{Beijing}
  \country{China}
}

\author{Lin Gan}
\affiliation{%
  \institution{Tsinghua University}
  \city{Beijing}
  \country{China}
}

\author{Guangwen Yang}
\affiliation{%
  \institution{Tsinghua University}
  \city{Beijing}
  \country{China}
}

\author{Zhiyuan Liu}
\affiliation{%
  \institution{Tsinghua University}
  \city{Beijing}
  \country{China}
}

\renewcommand{\shortauthors}{Chen et al.}

\begin{abstract}
Industrial and scientific computing rests on a small set of core kernels. The
stencil is among the most widely used of them, and it sits inside weather and
climate models, seismic imaging, fluid dynamics, and image processing.

No single stencil implementation is fastest. The optimal kernel changes
qualitatively with the stencil's shape, the grid's shape, the precision, and
the host application. For two decades the field has answered with
\emph{general} methods: DSLs, code generators, and autotuners. Specialized
solutions were too expensive to build per case, so all three reuse one
human-authored recipe across cases. That reuse costs performance. We call
the cost the generality tax.

We show that this premise no longer holds. Code-synthesis agents now build
a correct, specialized solution per case at a cost that is no longer
prohibitive. \sysname{} automates this at the operator level. A \kagent{}
synthesizes CUDA and forges a per-configuration map of specialized
operators, removing the generality tax case by case. The map beats the
strongest publicly available baseline for each case in 37 of 37 cases on an
A100: a geometric mean of \kernelgm{} against same-precision f32 baselines
and \kernelfphgm{} for fp16, each reported under its own precision so no
mixed-precision saving inflates the headline.

The same change reaches end-to-end application performance. A generic
operator library is, by construction, a compromise. It is tuned once for its
own general case and then reused across applications, so its shapes, layouts,
and launch boundaries cannot be optimal for any particular one. Using such a
library is the application-level form of the generality tax. An
\aagent{} forges a specialized solution per application
instead, locating hotspots, rewriting application structure, and validating
and integrating each change. Across \numapps{} real industrial and scientific
codes the end-to-end median speedup is \medspeedup{} against each
application's own GPU baseline. To our knowledge this is the first
demonstration that per-case synthesis carries from a kernel library to
complete applications at this breadth, and it is evidence that reuse is no
longer the default in a domain that has been built on it for two decades.
\end{abstract}



\maketitle

\section{Introduction}
\label{sec:intro}
\label{sec:intro:setting}

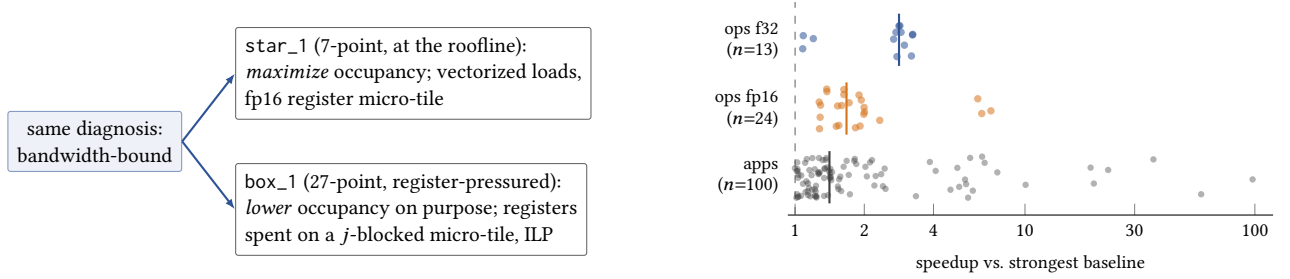
\begin{figure*}[t]
  \centering
  \begin{minipage}[c]{0.52\linewidth}
    \centering
    \begin{tikzpicture}[
        font=\footnotesize,
        box/.style={draw=black!60, rounded corners=1pt, align=left,
          inner sep=3.5pt, fill=white},
        >={Latex[length=4pt]},
      ]
      \node[box, fill=fsblue!8, draw=fsblue!60, align=center] (diag)
        {same diagnosis:\\bandwidth-bound};
      \node[box, right=7mm of diag.east, anchor=west, yshift=9mm] (star)
        {\texttt{star\_1} (7-point, at the roofline):\\
         \emph{maximize} occupancy; vectorized loads,\\
         fp16 register micro-tile};
      \node[box, right=7mm of diag.east, anchor=west, yshift=-9mm] (boxk)
        {\texttt{box\_1} (27-point, register-pressured):\\
         \emph{lower} occupancy on purpose; registers\\
         spent on a $j$-blocked micro-tile, ILP};
      \draw[->, fsblue, thick] (diag.east) -- (star.west);
      \draw[->, fsblue, thick] (diag.east) -- (boxk.west);
    \end{tikzpicture}
  \end{minipage}\hfill
  \begin{minipage}[c]{0.45\linewidth}
    \centering
    \begin{tikzpicture}
      \begin{axis}[
          width=\linewidth, height=44mm,
          xmode=log, log ticks with fixed point,
          xmin=0.93, xmax=120,
          ymin=-0.55, ymax=2.55,
          xtick={1,2,4,10,30,100},
          ytick={2,1,0},
          yticklabels={ops f32\\($n{=}13$),
                       ops fp16\\($n{=}24$),
                       apps\\($n{=}\numapps$)},
          yticklabel style={font=\scriptsize, align=right},
          xlabel={speedup vs.\ strongest baseline},
          xlabel style={font=\scriptsize},
          tick label style={font=\scriptsize},
          ytick style={draw=none},
          axis x line*=bottom, axis y line*=left,
          y axis line style={draw=none},
        ]
        \draw[dashed, black!60] (axis cs:1,-0.55) -- (axis cs:1,2.55);
        \addplot[only marks, mark=*, mark size=1.4pt, fsblue,
          fill opacity=0.55, draw opacity=0]
          table {figures/data/fig1_strip_ops_f32.tsv};
        \addplot[only marks, mark=*, mark size=1.4pt, fsorange,
          fill opacity=0.55, draw opacity=0]
          table {figures/data/fig1_strip_ops_fp16.tsv};
        \addplot[only marks, mark=*, mark size=1.2pt, black!75,
          fill opacity=0.40, draw opacity=0]
          table {figures/data/fig1_strip_apps.tsv};
        \draw[fsblue, thick]   (axis cs:2.832,1.62) -- (axis cs:2.832,2.38);
        \draw[fsorange, thick] (axis cs:1.674,0.62) -- (axis cs:1.674,1.38);
        \draw[black!75, thick] (axis cs:1.411,-0.38) -- (axis cs:1.411,0.38);
      \end{axis}
    \end{tikzpicture}
  \end{minipage}
  \caption{\sysname{} in one figure. \textbf{Left:} the same coarse diagnosis
    drives two stencils to \emph{opposite} forged structures
    (\S\ref{sec:kernel:techniques}), so any single generic policy pays the
    generality tax on the case it guessed wrong. \textbf{Right:} speedup
    distributions. Every forged operator beats the strongest public per-case
    baseline: 37 wins, 0 losses (kernel microbenchmarks, A100; f32 geomean
    \kernelgm{}, fp16 \kernelfphgm{}, reported separately by precision).
    End-to-end,
    the median across \numapps{} real applications is \medspeedup{}, measured
    by the application gate (\S\ref{sec:harness}); per-case detail in
    \S\ref{sec:experiments}. Dashed line is parity; the tick in each lane
    is that lane's median.}
  \label{fig:headline}
\end{figure*}

Stencil computation underlies a large fraction of industrial and scientific
computing, from computational fluid dynamics to climate codes. A stencil
repeatedly updates a grid from a fixed neighborhood of each element. These
kernels are memory-bound, and approaching the hardware limit is among the
most expert-intensive tasks in high-performance computing. The optimization
that reaches peak performance changes qualitatively with the exact stencil,
grid shape, precision, and target architecture (Figure~\ref{fig:headline},
left).

\label{sec:intro:tension}%
The dominant response over two decades has been to build \emph{general} methods
that cover many stencils, shapes, and applications from one human-authored
recipe. The systems are stencil
DSLs~\cite{ragankelley2013halide, luporini2020devito,
tang2011pochoir}, polyhedral and GPU code
generators~\cite{matsumura2020an5d, zhao2019bricks, holewinski2012stencil,
rawat2018stencilgen}, temporal-blocking
frameworks~\cite{zhang2023ebisu, you2021drstencil, nguyen2010blocking}, and
autotuners~\cite{datta2008stencil, ansel2014opentuner,
zheng2020ansor}. Such
systems do emit a per-case kernel, by compiling a schedule or searching a
parameter space. But the recipe, and the space it can reach, are fixed in
advance by a human. Specialized solutions were too expensive to build per
case, so all of these systems reuse one human-authored recipe. That reuse
costs performance, and we call the cost the generality tax.

This premise no longer holds. Code-synthesis agents produce a
correct, specialized
implementation from scratch at a marginal cost that is no longer
prohibitive~\cite{ouyang2025kernelbench, lange2025cudaengineer,
deepmind2025alphaevolve}. The basis for the generality tax therefore
disappears. Forging a fresh solution per case, and removing the tax case by
case, becomes practical.
We call this paradigm \forge{}, following the ForgeTrain line of work, which
develops it for deep-learning training frameworks~\cite{forgetrain_2026},
and we apply it to stencil computation in this paper.

\label{sec:intro:vehicle}%
\sysname{} begins at the operator level. For each concrete combination of
stencil type, grid shape, and precision, the \kagent{} writes a separate CUDA
kernel, instead of retuning one recipe
that spans cases. What keeps this from degenerating into independent one-off
kernels is a persistent operator map. Each accepted kernel contributes a
cell that binds a bottleneck diagnosis to the code that passed the gates, and
each rejected direction is recorded with the measurement that refuted it. A
new case therefore starts from the accumulated experience of every case
before it: it takes a neighboring cell's diagnosis---not its code---and
re-derives the structure for its own case rather than beginning from
scratch (\S\ref{sec:kernel:memory}), which is what carries generalization
across the map. A kernel is admitted only when its own gate measures it
correct and faster than the per-case public-SOTA upper bound. The map
beats that bound for each case in 37 of 37 cases.

The same applies one level up, to whole applications
(Figure~\ref{fig:headline}, right). A generic operator library is, by
construction, a compromise. It is tuned once for its own general case and
then reused across applications, so its shapes, layouts, and launch
boundaries cannot be optimal for any particular one. Using it is the
application-level form of the generality tax. An \aagent{} forges a
specialized solution per
application instead. It
locates the performance-critical region and forges a per-app
optimization, usually a structural rewrite, then validates and integrates
that change against the application's own tests and
timing, across \numapps{}+ real industrial and scientific codes, with an
end-to-end median speedup of \medspeedup{}.

Each level is checked by its own gate, and the two bodies of
evidence stay strictly separate. Operator-level claims rest only on kernel
microbenchmarks, application-level claims only on the application gate, and
neither endorses the other. We contribute an autonomous loop that designs,
measures, and deploys per-case specializations at both levels, and a
demonstration that this reaches end-to-end application performance.

\label{sec:intro:amenable}%
Forge Engineering has no workload-level admission test: the same hypothesis--measure--revise
loop applies wherever specialization can pay off in high-performance computing. We choose stencil
as the demonstration domain because its instances are clearly specified, per-case SOTA baselines
are abundant, correctness is easy to check, and performance can be measured separately at the kernel
and application levels. The dual-agent architecture itself is general: its
collaboration mirrors a standard high-performance-computing engineering
workflow. The agents form a hypothesis, implement an optimization, measure
the result, and iterate on the feedback. Nothing stencil-specific lives in
the loop's rules; that knowledge sits entirely in the operator map
(\S\ref{sec:app:twolayer}), and the same loop already forges the non-stencil
hotspots it meets inside real applications (\S\ref{sec:taxonomy:levers}).
Carrying the architecture to FFT, SpMV, tensor contraction, or GEMM fusion
is therefore a matter of growing a new matrix, not of redesigning the loop.

\medskip
\noindent\textbf{Contributions.} We make three contributions.

\begin{itemize}
  \item \textbf{Kernels that beat per-case SOTA, 37 cases of 37.} We design a configuration-indexed matrix of specialized
    operators, which is the \kagent{}'s own knowledge base: application-level forging
    invokes the \kagent{} on the hot-operator case, and the match decision happens
    inside it. Across mainstream stencil types, shapes, and f16/f32/f64 precisions, the
    library wins 37 of 37 cases against per-case public-SOTA upper bounds
    (same-precision f32 geometric mean \kernelgm{} on A100).
  \item \textbf{End-to-end speedups on \numapps{}+ real applications.} We design a staged workflow that performs framework-level
    optimization before operator-level forging: it first rewrites scheduling, fusion, and host--device
    traffic, then invokes the \kagent{} for the hot operator, and finally validates and
    integrates the change.
    The workflow covers \numapps{}+ real industrial and scientific codes and achieves an end-to-end
    median speedup of \medspeedup{} against same-architecture GPU baselines.
  \item \textbf{A harness that enforces the protocol.} Separate operator and application gates
    verify the two levels through executable checks on agent-read-only measurement surfaces. An
    overstated speedup becomes a system-level error, and the gates retract the framework's own
    numbers when they fail their checks. An ablation shows that removing any
    single requirement distorts a real measurement (\S\ref{sec:harness:ablation}).
\end{itemize}
\section{Background \& Motivation}
\label{sec:background}

\subsection{Stencils, their variety, and the bandwidth limit}
\label{sec:background:roofline}

A stencil computation repeatedly updates each element of a regular grid from a
fixed set of neighboring elements. Let $u^{t}$ be the grid state at time step
$t$ over a $d$-dimensional index domain $\Omega \subseteq \mathbb{Z}^{d}$. A
stencil is fixed by a finite \emph{offset set} $S = \{\delta_{1},\dots,\delta_{k}\}
\subset \mathbb{Z}^{d}$, the pattern of neighbors each point reads. It is
paired with an update rule
\[
  u^{t+1}(x) \;=\; \Phi\!\left(\{\, u^{t}(x+\delta) : \delta \in S \,\}\right),
  \qquad x \in \Omega ,
\]
which is applied to every point of the grid and iterated over $t$.
In the common linear,
constant-coefficient case $\Phi$ collapses to a weighted sum
$u^{t+1}(x) = \sum_{\delta \in S} c_{\delta}\, u^{t}(x+\delta)$, so a stencil is,
concretely, the pair (offset geometry $S$, coefficients $c_{\delta}$) driving a
time-stepped sweep. It is the core of a large class of scientific and industrial
codes: fluid dynamics, seismic and electromagnetic modeling, climate and weather,
molecular dynamics, and the discretized PDE solvers embedded in many simulation
tools.

Instances of this pattern differ along three axes: the stencil's type, the
run's shape, and the working precision. The \emph{type} is the geometry of
$S$. A \emph{star} pattern reads neighbors only along the axes; a \emph{box}
pattern reads the full surrounding cube. A higher order
$r=\max_{\delta\in S}\lVert\delta\rVert_{\infty}$ or dimension $d$ means each
point reads more neighbors, so it does more arithmetic and touches a larger
region of the grid. If the coefficients $c_{\delta}$ are constant, they can
live in registers; if they vary in space, they are one more array to read
from memory. The \emph{shape} is the size and aspect ratio of $\Omega$: the
same stencil may run on a cube or on a thin, elongated slab. The
\emph{precision} (f16/f32/f64) sets how many bytes each grid value occupies.
Together, the three axes decide what data each point must read and how many
bytes the run moves. The rest of this subsection shows that this is exactly
what sets performance.

Stencils are arithmetic-light. Arithmetic intensity makes that precise: it is
the number of arithmetic operations performed per byte moved from DRAM. One
seven-point star update in f32 reads seven values and writes one, 32 bytes, and
performs seven multiply-adds, 14 FLOP. That is roughly 0.4 FLOP per byte.
Modern GPUs can perform on the order of ten FLOP per byte they can move, so
this kernel sits far on the memory side of the balance. Its arithmetic is
nearly free; what costs time is moving data. Raising the intensity is therefore
the only way to speed it up. On-chip reuse is what raises it: a kernel that
keeps a block of the grid resident reuses each byte it loads across many
updates instead of reading it once per update. This reuse is what can push a
kernel into the compute-bound region, and whether it does depends on the
stencil's type and the run's shape. Those decide how much of the grid a block
must keep resident for the reuse to pay. For most instances in our set none of
it is enough: the library's measured operators reach \rooflinepeak{} of peak
HBM bandwidth (median \rooflinemedian{}, \S\ref{sec:exp:operator}), which
confirms the regime rather than assuming it.

Peak performance is therefore a question of memory traffic, and the levers all
act on it. Read less of what has already been read, reuse across registers and
shared memory, arrange accesses so that neighboring threads read neighboring
addresses, and keep occupancy high enough to hide latency. The levers trade
against one another, and type and shape decide the terms of the trade. A larger
tile, for example, buys more reuse but spends registers or shared memory, which
lowers occupancy, and the block size that balances the two moves with the
instance.

Because the position on that roofline moves with the instance, so does the
value of each lever. Consider shape alone, with stencil type, precision, and
architecture held fixed. Cubes and anisotropic slabs of the same seven-point
star route to different kernels and different launch geometries. A kernel
hand-tuned to the hardware limit at one $(S, \text{shape}, \text{precision},
\text{architecture})$ point rarely stays optimal at the next. Type and
precision move the arithmetic intensity itself. Raising the order $r$ or the
dimension $d$ increases both the arithmetic per point and the reuse available.
Lowering precision moves the same computation toward the compute-bound side by
shrinking the bytes per point. The optimum is
therefore a property of the instance, not of the stencil family, which is what
makes per-case specialization both necessary and worth automating.

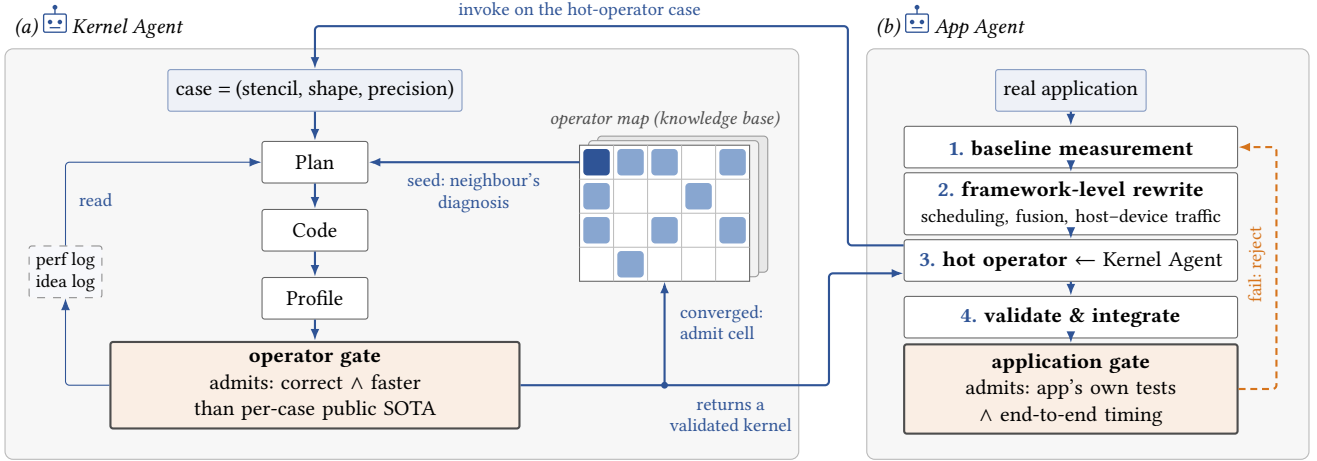
\begin{figure*}[t]
  \centering
          \begin{tikzpicture}[
      font=\footnotesize,
      box/.style={draw=black!60, rounded corners=1pt, align=center,
        inner sep=3pt, minimum height=5.5mm, fill=white},
      stage/.style={box, minimum width=14mm},
      chip/.style={box, fill=fsblue!8, draw=fsblue!60},
      gatebox/.style={box, draw=black!70, thick, fill=fsorange!12},
      logbox/.style={draw=black!45, dashed, rounded corners=1pt, inner sep=2.5pt,
        align=center, fill=fsblue!4, font=\scriptsize},
      panel/.style={draw=black!28, rounded corners=3pt, fill=black!3},
      ptitle/.style={font=\footnotesize\itshape, anchor=south west},
      fwd/.style={-{Latex[length=4.5pt]}, fsblue, thick},
      fail/.style={-{Latex[length=4.5pt]}, fsorange, thick,
        dash pattern=on 2.6pt off 1.8pt},
      lbl/.style={font=\scriptsize, inner sep=1.5pt},
      rounded corners=1.5pt,
    ]

    \tikzset{robot/.pic={
      \draw[fsblue, fill=white, rounded corners=0.6pt, line width=0.5pt]
        (-1.5mm,-1.2mm) rectangle (1.5mm,1.2mm);
      \fill[fsblue] (-0.65mm,0.15mm) circle (0.3mm);
      \fill[fsblue] (0.65mm,0.15mm) circle (0.3mm);
      \draw[fsblue, line width=0.5pt] (-0.7mm,-0.6mm) -- (0.7mm,-0.6mm);
      \draw[fsblue, line width=0.5pt] (0,1.2mm) -- (0,1.9mm);
      \fill[fsblue] (0,2.1mm) circle (0.32mm);
    }}

    \node[chip] (case) at (45mm,-12mm) {case $=$ (stencil, shape, precision)};
    \node[stage] (kp) at (45mm,-21.5mm) {Plan};
    \node[stage] (kc) at (45mm,-30.5mm) {Code};
    \node[stage] (kf) at (45mm,-39.5mm) {Profile};
    \node[gatebox, text width=52mm] (kgate) at (45mm,-51mm)
      {\textbf{operator gate}\\
       admits: correct $\wedge$ faster than per-case public SOTA};
    \node[logbox] (mem) at (12mm,-36mm) {perf log\\idea log};

    \draw[fwd] (case) -- (kp);
    \draw[fwd] (kp) -- (kc);
    \draw[fwd] (kc) -- (kf);
    \draw[fwd] (kf) -- (kf |- kgate.north);
    \draw[fwd, thin] (kgate.west) -- (12mm,-51mm) -- (mem.south);
    \draw[fwd, thin] (mem.north) -- (12mm,-21.5mm) -- (kp.west);
    \node[lbl, text=fsblue, anchor=west] at (13.2mm,-26.5mm) {read};

    \begin{scope}
      \draw[fill=black!9, draw=black!40] (82.4mm,-36.05mm) rectangle +(22.5mm,18mm);
      \draw[fill=black!5, draw=black!40] (81.2mm,-36.65mm) rectangle +(22.5mm,18mm);
      \draw[fill=white, draw=black!60, sharp corners]
        (80mm,-37.25mm) rectangle +(22.5mm,18mm);
      \foreach \c/\r in {1/3,2/3,4/3,0/2,3/2,0/1,2/1,4/1,1/0}
        \fill[fscell] (\c*4.5mm+80.5mm,-37.25mm+\r*4.5mm+0.5mm) rectangle +(3.5mm,3.5mm);
      \fill[fsblue] (80.5mm,-23.25mm) rectangle +(3.5mm,3.5mm);
      \foreach \x in {1,...,4} \draw[black!25] (\x*4.5mm+80mm,-37.25mm) -- +(0,18mm);
      \foreach \y in {1,...,3} \draw[black!25] (80mm,-37.25mm+\y*4.5mm) -- +(22.5mm,0);
    \end{scope}
    \node[lbl, font=\scriptsize\itshape, text=black!70, anchor=south]
      at (91.25mm,-17.6mm) {operator map (knowledge base)};

    \draw[fwd] (80mm,-21.5mm) -- (kp.east);
    \node[lbl, align=center, text=fsblue, anchor=north] at (66mm,-22.4mm)
      {seed: neighbour's\\diagnosis};
    \draw[fwd] (91.25mm,-51mm) -- (91.25mm,-37.25mm);
    \fill[fsblue] (91.25mm,-51mm) circle (0.45mm);
    \node[lbl, align=left, text=fsblue, anchor=west] at (92.7mm,-42.7mm)
      {converged:\\admit cell};

    \node[chip] (app) at (145mm,-12mm) {real application};
    \node[stage, text width=42mm] (m0) at (145mm,-19.5mm)
      {\textbf{\textcolor{fsblue}{1.}\ baseline measurement}};
    \node[stage, text width=42mm] (m1) at (145mm,-27mm)
      {\textbf{\textcolor{fsblue}{2.}\ framework-level rewrite}\\
       {\scriptsize scheduling, fusion, host--device traffic}};
    \node[stage, text width=42mm] (m2) at (145mm,-34.5mm)
      {\textbf{\textcolor{fsblue}{3.}\ hot operator} $\leftarrow$ \kagent{}};
    \node[stage, text width=42mm] (m3) at (145mm,-42mm)
      {\textbf{\textcolor{fsblue}{4.}\ validate \& integrate}};
    \node[gatebox, text width=42mm] (agate) at (145mm,-51.5mm)
      {\textbf{application gate}\\
       admits: app's own tests $\wedge$ end-to-end timing};
    \draw[fwd] (app) -- (m0);
    \draw[fwd] (m0) -- (m1);
    \draw[fwd] (m1) -- (m2);
    \draw[fwd] (m2) -- (m3);
    \draw[fwd] (m3) -- (agate);
    \draw[fail] (agate.east) -- ++(5mm,0) |- (m0.east);
    \node[lbl, rotate=90, text=fsorange, inner sep=1pt]
      at ([xshift=2.5mm]m1.east |- 0,-35mm) {fail: reject};

    \begin{scope}[on background layer]
      \draw[panel] (4mm,-60.8mm) rectangle (109mm,-6.5mm);
      \draw[panel] (118mm,-60.8mm) rectangle (176.5mm,-6.5mm);
    \end{scope}
    \node[ptitle] at (4mm,-6.3mm) {(a)~\tikz\pic{robot};~~\kagent{}};
    \node[ptitle] at (118mm,-6.3mm) {(b)~\tikz\pic{robot};~~\aagent{}};

    \draw[fwd] ([yshift=2mm]m2.west) -- (115.5mm,-32.5mm)
      -- (115.5mm,-3.5mm) -- (45mm,-3.5mm) -- (case.north);
    \node[lbl, text=fsblue, anchor=south] at (80mm,-3.2mm)
      {invoke on the hot-operator case};
    \draw[fwd] (kgate.east) -- (113mm,-51mm)
      -- (113mm,-36.2mm) -- ([yshift=-1.7mm]m2.west);
    \node[lbl, align=center, text=fsblue, anchor=north] at (100mm,-52.2mm)
      {returns a\\validated kernel};
  \end{tikzpicture}
  \caption{\sysname{} at a glance. \textbf{(a)}~The \kagent{} forges one
    kernel per (stencil, shape, precision) case. Its operator map is the
    knowledge base: each cell binds a bottleneck diagnosis to the code that
    cleared the operator gate, which admits a kernel only when it is
    correct and faster than the per-case public-SOTA upper bound. The map
    is consulted once, when a new case is planned. The new loop starts from a
    neighbor's diagnosis, not its code (\S\ref{sec:kernel:memory}). A converged
    loop contributes its cell back. Within a loop, a round's one change is
    committed if it
    measures faster and reverted otherwise; committed or not, the round is
    recorded: the perf log and the idea log are what each round writes and
    the next round reads. \textbf{(b)}~The \aagent{} measures the application's
    baseline first, rewrites its framework next, and only then obtains the hot
    operator: it invokes the \kagent{} and takes back a validated kernel,
    which either answers from the map or forges the missing cell. It
    validates and integrates the result last
    (\S\ref{sec:app}, which resolves these
    four steps into the M1--M4 milestones). Each gate judges only its own
    level.}
  \label{fig:arch}
\end{figure*}

\subsection{Generic stencil methods and the generality tax}
\label{sec:background:prior}

The field answered this per-case pressure with \emph{general} methods along
three routes, which differ in what they reuse. \emph{Parallelizing compilers
and code generators} reuse transformation rules: compilation frameworks
(Pochoir~\cite{tang2011pochoir}, PATUS~\cite{christen2011patus},
Physis~\cite{maruyama2011physis}, YASK~\cite{yount2016yask}), the polyhedral
generators AN5D~\cite{matsumura2020an5d} and Bricks~\cite{zhao2019bricks},
and single-lever generators for on-chip reuse and register pressure
(Overtile~\cite{holewinski2012stencil},
StencilGen~\cite{rawat2018stencilgen}, register-level
optimization~\cite{rawat2018register}, Lift~\cite{hagedorn2018lift}).
\emph{Autotuning and tensor compilers} reuse search templates:
stencil autotuning~\cite{datta2008stencil},
configuration search~\cite{ansel2014opentuner}, and schedule search under
a learned cost model in TVM~\cite{chen2018tvm} and
Ansor~\cite{zheng2020ansor}. \emph{Hand-optimized
implementations and DSLs} reuse the implementation itself, and are the
stencil counterpart of the
vendor-library route. Halide~\cite{ragankelley2013halide} and
Devito~\cite{luporini2020devito} separate schedule from algorithm to lower
the authoring bar, and hand-optimized lines set the standard per family:
temporal blocking (time skewing and its 3.5-D
refinement~\cite{nguyen2010blocking}, EBISU~\cite{zhang2023ebisu},
DRStencil~\cite{you2021drstencil}) and matrix-unit remapping
(ConvStencil~\cite{li2024convstencil},
FlashFFTStencil~\cite{han2025flashfftstencil}).

Generality is bought with reuse, and the generality tax is what the purchase
costs. To span many cases, a route must fix one thing in advance and reuse it
everywhere, so each case it was not built for pays for the difference. What
gets reused differs by route, and so does the form the tax takes. A compiler
reuses one set of transformation rules, so a case that needs cross-kernel
fusion or a memory layout rebuilt for its setting can only be served by a
schedule those rules already express. An autotuner reuses one search template,
so a case receives a better parameter setting inside that template's form and
never a different form. A library or DSL reuses one implementation with its
built-in variants, so a case outside those variants runs on design decisions
made for the cases the implementation does serve. In each route the reuse is
what makes the wide coverage possible, and it is also what the off-target case
pays for. Thin the abstraction layer, the configuration surface, and the code
paths written for other cases, and the coverage narrows while the cost stays.

A lever's value is a property of the instance, so any fixed set of rules or
templates pays the tax wherever the roofline argument above places the optimum
elsewhere. The tax has also gone unmeasured: these systems are
validated almost exclusively on microbenchmarks, each against the defaults,
and are almost never compared head-to-head on the same cases.

The systems above play two roles in this paper. They serve as per-case
baselines: any of them can stand in for the SOTA upper bound we measure
against, and several do, so they define what specialized stencil performance
currently means. They also mark where we differ, and the difference is the
artifact. A generic method produces one reusable implementation, and its
coverage is bounded by the shapes and interfaces it anticipated. \sysname{}
produces a repeatable forging process, and its coverage grows with each case it
is given. Stencil is our demonstration domain because it offers clearly defined
instances, strong per-case baselines, and a demanding setting for controlled
measurement.

\subsection{Why per-case specialization becomes practical now}
\label{sec:background:llm}
\label{sec:background:forge}

Because the three routes reuse rules, templates, or implementations, the tax
has persisted for an economic reason rather than a technical one:
implementations were too expensive to build per case. Three lines of recent work change that cost. Program search now finds faster
low-level algorithms directly~\cite{fawzi2022alphatensor,
mankowitz2023alphadev, real2020automlzero, chen2023symbolic,
mirhoseini2021graphplacement}. Autonomous agents supply the loop machinery to
run such synthesis without a human in each round~\cite{yao2023react,
shinn2023reflexion, lu2024aiscientist, jimenez2024swebench}. Closest to our
setting, agents now synthesize performance-critical GPU code
itself~\cite{lange2025cudaengineer, deepmind2025alphaevolve,
ouyang2025kernelbench}. An implementation per case is now cheap enough to
produce that reuse stops being the rational default. The cross-case
abstractions, configuration surfaces, and code paths written for other cases
were there to recover the cost of that implementation, and with the cost gone
there is nothing left to recover. Per-case forging can remove the tax at its
source rather than optimize within its constraints.

Cheap synthesis alone, however, is not performance engineering. Performance
engineering is iterative: form a hypothesis, implement it, measure correctness
and speed, and revise from the feedback. Recent LLM-driven operator optimizers
already couple synthesis to hardware feedback or iterative coding
agents~\cite{deng2025autostencil, lange2025cudaengineer,
deepmind2025alphaevolve}, matching this workflow. The natural form is
therefore an agent loop. The loop holds the state between rounds, measures the
result of each round, and repeats until a candidate passes its gates. Unlike
the three routes above, it carries no fixed form. What it knows is the record
of cases it has already forged, so that record grows with use instead of being
fixed in advance.

This line of work leaves three gaps, and they define what \sysname{} has to
supply. First, it has barely reached stencil. Kernel synthesis focuses on GEMM,
attention, and scheduling, and the little stencil-specific work trains on a
corpus of existing implementations~\cite{deng2025autostencil}, which makes it
depend on that corpus. Second, it treats each kernel as a one-off artifact
rather than specialization as a repeatable engineering practice. Third, trust
remains unresolved. Reward hacking is documented, and removing contaminated
tasks drops the reported KernelBench aggregate from $3.13\times$ to
$1.49\times$~\cite{lange2025cudaengineer}. A result is only as trustworthy as
the harness that measures it.
\section{System Overview}
\label{sec:overview}

\sysname{} realizes \forge{} for stencil computation through two cooperating
agents\label{sec:overview:agents}. Figure~\ref{fig:arch} shows the whole
system. Panel~(a) is the \kagent{}, which works on operators. Panel~(b) is the
\aagent{}, which works on whole applications. The two panels are joined by a
single arrow: the \aagent{} invokes the \kagent{} and gets back a validated
kernel. This section reads the figure panel by panel.

\textbf{Panel (a): the \kagent{}.} The \kagent{} forges one kernel per
(stencil type, grid shape, precision) case. The forged kernels form the
operator map, which is also the loop's knowledge base: a cell binds a
bottleneck diagnosis to the code that cleared the operator gate. The loop
runs in three stages. It plans the next change, writes it into the kernel, and
profiles the result. The operator gate then admits the change only if it is
correct and faster than the per-case public-SOTA upper bound. A round is
committed if it passes and reverted otherwise, and either way it is recorded:
the perf log and the idea log are what a round writes and the next round reads.
Section~\ref{sec:kernel} describes the loop, the map, and the kernels it
forges.

\textbf{Panel (b): the \aagent{}.} The \aagent{} starts from a real
application rather than from a kernel. It measures the application's baseline
first, so that every later speedup is reported against the same reference. It
then rewrites the application's framework
structure, because scheduling, fusion, and host--device traffic bound many
applications before any single operator does. Only after that does it need the hot
operator, and it obtains that operator by invoking the \kagent{}. Finally it
validates and integrates the result.
Section~\ref{sec:app} defines the \aagent{}'s milestones, exit states, and
admission gates.\label{sec:overview:forging}

\textbf{Shared mechanism.} Both agents only propose changes. Each level runs one gate,
and that gate admits a change only after the \harness{} has measured it on
real hardware. That separation is what makes
autonomous forging checkable: the loop that writes the code is never the judge
of it. Each gate judges only its own level. The operator gate never sees an
application's end-to-end time, and the application gate never sees a kernel in
isolation. The dependency between
the two agents is one-way as well. The \aagent{} invokes the \kagent{}, but the
matrix is never reached from outside the \kagent{}, so no application can
author what the next case is seeded with.
\section{\kagent{}: Forging Specialized Operators}
\label{sec:kernel}

The \kagent{} is where \forge{} happens at the operator level. It takes one
concrete combination at a time: a stencil type, a grid shape, and a precision.
For that combination it writes a specialized CUDA kernel, and it keeps the
kernel only if it measures faster and correct. This section describes the loop,
then the two-layer memory the loop builds up in the operator map, and
finally the structures it forges. Those structures also show why the loop is
doing code synthesis rather than autotuning or template instantiation.

\begin{figure}[t]
  \centering
  \includegraphics[width=\linewidth]{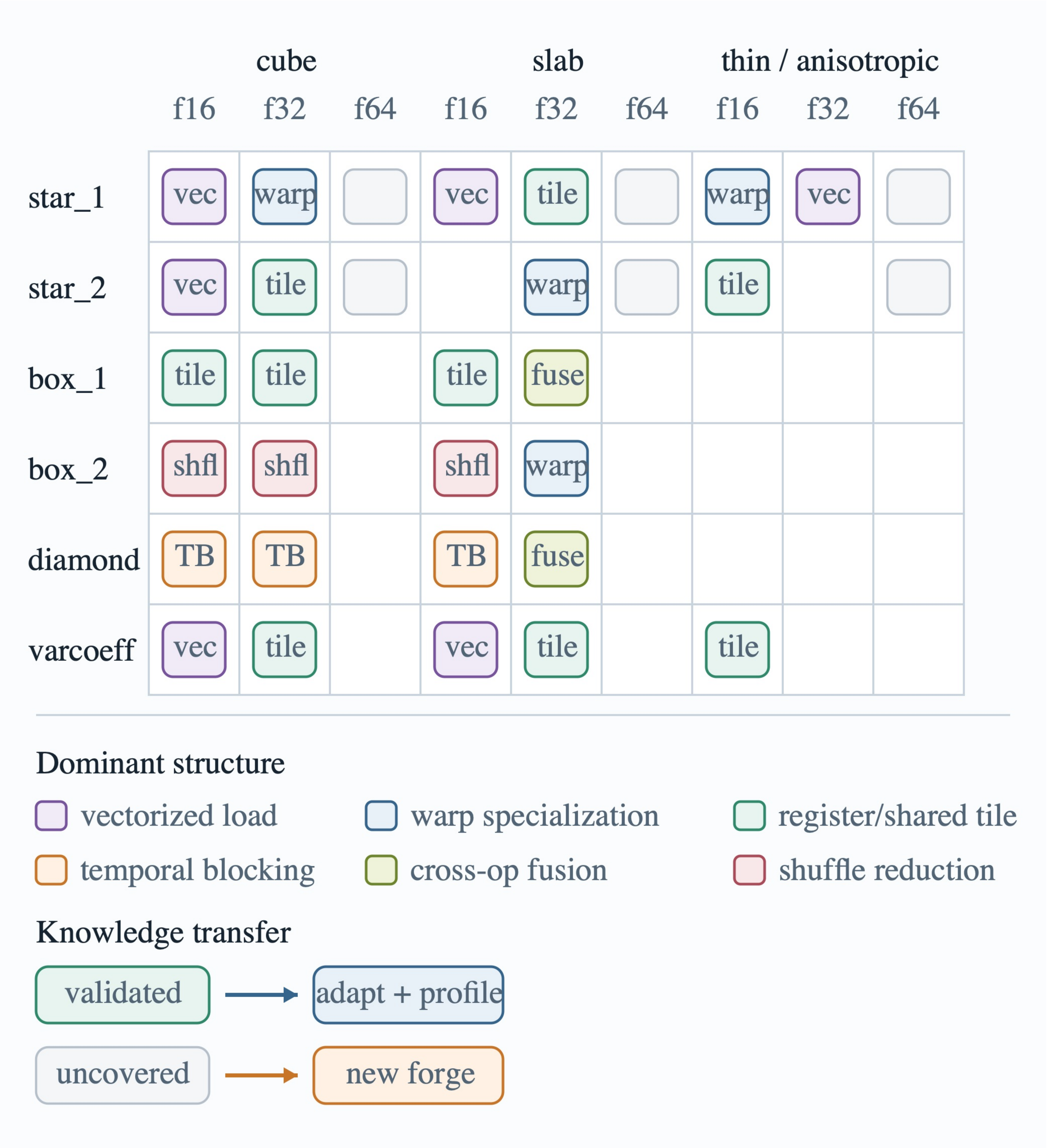}
  \caption{The operator map indexes the dominant structure by stencil,
    shape, and precision. A validated cell seeds adaptation and profiling for
    new cases; an uncovered cell triggers a fresh forge.}
  \label{fig:operator-map}
\end{figure}

\subsection{The code-synthesis autoloop}
\label{sec:kernel:autoloop}
\textbf{One round.} The \kagent{} runs an autonomous loop, with no human inside
a round. Each round begins by reading five inputs: the rulebook, the
performance log, the idea log, the persisted baseline, and the last few
commit subjects. The first four carry what the loop has learned. The commit
subjects tell the round where it stands in the sequence of
accepted changes. The agent then proposes one change and edits only the CUDA
kernel source and its Python entry point. It measures the affected case first,
and the whole matrix after that. The operator gate applies its two rules: correctness
against an independent oracle, and no regression against the accepted cases.
The change is committed if both pass and reverted otherwise. Either way the
round is recorded. The agent writes the outcome into the
performance log, and the next concrete idea into the idea log, so a reverted
round still informs the rounds that follow. Then it
stops. It is
told explicitly that no user is present, so a question would spend a round
without producing a measurement.

\textbf{Context management.} Each round starts from an empty context. The
agent is a fresh process, and nothing survives from the previous round except
what was written to disk, so a round's starting state is the source tree
itself. Carrying hundreds of rounds of history would inflate the context
until attention no longer stays on the case at hand. Most of that history is
also redundant: the kernel source already describes where the
search has arrived. The tree has one blind spot. A reverted change leaves
no trace, so the code cannot show what was already tried and failed. The
logs cover exactly that gap, which is why a round opens by reading them. Log
entries must
stay factual. Each one ties a change to its measurement and a claim to its
profile, because every later round reads the log and would otherwise take an
unsupported opinion for an established result.

\textbf{One change per round.} The one-change rule keeps a round's outcome
readable. When the map regresses, the cause is that round's single edit,
which is reverted without any later bisection. The log entry names one
specific structure, precisely enough for a later round to act on it.

\textbf{Branching.} Some structures need several rounds to get right. Under a
per-round verdict they would be abandoned at their first failure. For that
reason, disruptive
rewrites and single-shape kernels are built on their own branches. A
disruptive rewrite is something like a wholesale fusion or a tensor-core
mapping. After five
rounds without progress, a branch is measured against the main line. It is
merged if faster, and dropped otherwise. Every accepted round is a commit, so
dropping a branch costs nothing that was already working.

\subsection{The map as the loop's memory}
\label{sec:kernel:memory}
\textbf{The map.} The forged kernels form an operator map. Each cell
carries the winning lever and the concrete code that won with it, so the map
serves as the loop's memory between rounds. It sits inside the forging loop
rather than in the harness, and that placement is deliberate. A cell binds a
bottleneck diagnosis to the code that cleared the gate, so only the loop can
reuse it, and the gate never reads it. This is where we part ways with
ForgeTrain, whose knowledge prior lives in the \harness{} so that no
agent-written artifact can influence admission. Here the cell that is consulted
is exactly the artifact that was admitted, and the gate constrains admission
independently of what the map holds.

\textbf{Positive layer.} The source tree is a library of worked examples. Each
cell couples a bottleneck diagnosis to code that cleared the gate. A new case
borrows a nearby cell's diagnosis, re-derives the structure that diagnosis
admits for its own case, and re-profiles the result. The kernel is kept only if
it passes. Borrowing transfers the diagnosis, not the code: what the loop takes
from the neighbor is which structure that bottleneck class admits, and it
re-derives the kernel for the new case from that starting point. The map is
consulted once, when the case is planned, and written once, when the loop for
it converges. Inside a loop the round's outcome goes to the logs below, not to
the map.

\textbf{Negative layer.} The logs record each rejected direction with its loss,
its profile-backed cause, and its stopping condition. One example: a register
micro-tile lost $1.4\%$ on a \texttt{star\_1} slab, because wave quantization
rather than memory-level parallelism limited the scheduler. The log therefore
rules out fewer, fatter blocks and points the next round toward thinner ones.

\textbf{Control.} Repeated dead ends become rules that constrain later rounds.
After three consecutive non-improving rounds, the loop must search outside its
history. Both layers are plain flat files, reread at the start of each round.
There is no retrieval index.

\subsection{How forging removes the generality tax}
\label{sec:kernel:techniques}
\textbf{The mechanism.} The background argued that the generality tax is
structural. The forged library makes that argument concrete. The levers
the loop pulls are the standard repertoire for a bandwidth-bound kernel:
register coarsening, vectorized loads, shared-memory tiling, and occupancy
tuning. That repertoire is generic. What is specialized is the choice. The loop
profiles a case, finds the binding bottleneck, and pulls only the levers
that pay off for it.

\textbf{Opposite decisions.} Two bandwidth-bound kernels show what that means
in practice. The same coarse diagnosis drives them to \emph{opposite}
decisions. \texttt{star\_1} (7-point) already runs near the DRAM roofline, so
the loop pushes the only levers left: an fp16 register micro-tile, vectorized
loads, maximal occupancy, and a launch geometry split in two. One variant
serves cubes and one serves slabs. \texttt{box\_1} (27-point) is
bandwidth-bound too, but it runs under register pressure, so the loop
\emph{lowers} occupancy on purpose. It spends registers on
a $j$-blocked micro-tile and leans on instruction-level parallelism
instead. The same diagnosis leads to opposite occupancy decisions. A
generic recipe must commit to one policy in advance, and it pays the tax on
the kernel it guessed wrong. The loop instead forges for each kernel
the strategy its bottleneck demands.

\textbf{Real baselines.} No generic baseline wins across the full matrix.
The per-case winners are split among Halide, Devito, and EBISU.
Table~\ref{tab:baseline-identity} summarizes the 37 cases with a public-SOTA
baseline; the complete per-case mapping is in the supplementary material.

\begin{table}[t]
  \centering
  \caption{Per-case public-SOTA baseline ownership across 37 cases.}
  \label{tab:baseline-identity}
  \small
  \begin{tabular}{@{}lrrr@{}}
    \toprule
    & Halide & Devito & EBISU \\
    \midrule
    fp16 ($n{=}24$) & 21 & 3  & 0 \\
    fp32 ($n{=}13$) & 3  & 7  & 3 \\
    \bottomrule
  \end{tabular}
\end{table}

\textbf{A shifting axis.} In the next two cases the bottleneck itself moves,
so the optimization \emph{axis} changes. No retuning of a bandwidth-axis
recipe reaches them. \texttt{box\_2} (125-point) is the library's one
compute-bound corner. There the lever shifts to cutting computation with a
warp-level \texttt{shfl} reduction, and the loop also forged a tensor-core
mapping, measured it, and rejected it. The
\texttt{diamond} temporal-blocking family moves the axis the other way, to
temporal depth. Several time steps fuse into one launch. Intermediate time
planes stay on-chip, so register and shared-memory reuse
replaces per-step DRAM streaming. Measurement merged the 4- and 8-step
variants into a single fused kernel, where blind per-configuration
generation would have multiplied them. The same fusion idea extends to
\numfused{} kernels that fuse a stencil with the application's physics into
one launch. No lever is applied uniformly, and each appears only where a
measurement justified it.

\textbf{Beyond a fixed template.} Numeric sweeps are peeled off into
deterministic scripts, and the \kagent{} instead synthesizes the
concrete CUDA \emph{structure}.\label{sec:kernel:vs-autotuner} An equal-compute
ablation separates synthesis, search, and one-shot generation empirically:
neither search nor one-shot generation reaches what the loop forges
(\S\ref{sec:exp:synth-vs-search}).
\section{\aagent{}: Extending Specialization to Whole Applications}
\label{sec:app}

The \aagent{} carries \forge{} from the operator level to whole applications.
For each real codebase it forges an optimization specialized to that
application, validates it with the program's own tests and timing, and
integrates it. The central finding of this section is that, with few
exceptions, these wins are per-application structural
solutions\label{sec:app:insight}. They fuse launches, remove host-device
round-trips, fix a layout that defeats coalescing, or simplify an arithmetic
formulation. Drop-in replacement of a stencil kernel accounts for a small
minority, and an unmodified generic-library operator for none of the
\numapps{} wins (\S\ref{sec:taxonomy:levers}). Specialization therefore has
to be forged per application instead of built once and reused everywhere.

\subsection{States, milestones, and the Amdahl ceiling}
\label{sec:app:amdahl}
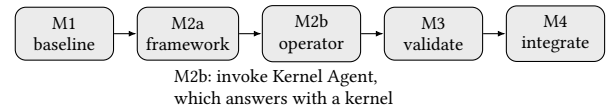
\begin{figure}[H]
  \centering
  \begin{tikzpicture}[
      font=\scriptsize,
      milestone/.style={draw, rounded corners, fill=black!8, minimum width=13mm,
        minimum height=7mm, align=center},
      >={Latex[length=3pt]},
    ]
    \node[milestone] (m1) {M1\\baseline};
    \node[milestone, right=3mm of m1] (m2a) {M2a\\framework};
    \node[milestone, right=3mm of m2a] (m2b) {M2b\\operator};
    \node[milestone, right=3mm of m2b] (m3) {M3\\validate};
    \node[milestone, right=3mm of m3] (m4) {M4\\integrate};
    \draw[->] (m1)--(m2a); \draw[->] (m2a)--(m2b);
    \draw[->] (m2b)--(m3); \draw[->] (m3)--(m4);
    \node[align=left] at (2.9,-0.7) {M2b: invoke \kagent{},\\which answers with a kernel};
  \end{tikzpicture}
  \caption{Application forging proceeds from baseline to framework optimization,
    obtaining the hot operator from the \kagent{}, validation, and integration.}
  \label{fig:app-milestones}
\end{figure}
An integration attempt has several honest outcomes besides success, and each
should be recorded. The \aagent{} therefore tracks every candidate with an
explicit state machine. A candidate starts in \texttt{PENDING} and leaves it
only by reaching one of five terminal states:
\begin{itemize}
  \item \texttt{OUT\_OF\_SCOPE}: no stencil-shaped hotspot, outside
    \forge{}'s premise.
  \item \texttt{BLOCKED}: no fair baseline can be established, for example
    because the application's own GPU baseline is unavailable.
  \item \texttt{RESEARCH}: a real opportunity exists, but a validated
    integration is out of reach this pass.
  \item \texttt{DROPPED}: forging and validation ran, but nothing cleared the
    gates.
  \item \texttt{INTEGRATED}: a change passed the correctness and timing
    gates and was merged.
\end{itemize}
The non-success states are first-class outcomes. Recording them keeps the
reported coverage honest.

The states are assigned at five milestones, so an integration can stop and
be classified at a well-defined point. \textbf{M1} locates the
performance-critical region and establishes the application's own GPU
baseline. \texttt{OUT\_OF\_SCOPE} and \texttt{BLOCKED} exit here.
\textbf{M2a} performs framework-level optimization. \textbf{M2b} invokes the
\kagent{} on the hot-operator case and takes back a validated kernel.
\textbf{M3} validates the change against the program's built-in correctness
check and timing. \textbf{M4} integrates it. Every gate measures against the
M1 baseline, the application's own GPU implementation. Correctness and speed
are independent gates. The application gate enforces both.

What is achievable end to end is bounded by Amdahl's law. If the targeted region
takes a fraction $f$ of end-to-end runtime and is accelerated by a factor $S$, the
whole-application speedup cannot exceed $1/(1 - f + f/S)$. Two consequences recur
through the results. First, a small stencil fraction $f_{\text{stencil}}$
caps the end-to-end gain however fast the operator becomes. Chasing operator
speed alone is therefore often the wrong target. Second, the largest
end-to-end wins tend to come from raising $f$ structurally, not from
shrinking a single kernel's time. The structural means are removing
redundant launches, host-device traffic, and recomputation. This ceiling
also sets the order in which M2a and M2b forge.

\subsection{Two-layer forging, framework before operators}
\label{sec:app:twolayer}
M2 consists of two milestones in a fixed order: framework-level optimization
first (M2a), operator-level optimization second (M2b). Framework-level
optimization targets the structure around the kernels, mainly scheduling,
operator fusion, and removing redundant launches, host-device round-trips,
and recomputation. Operator-level optimization targets the hot kernel
itself. The \kagent{} carries it out with the same forge loop, now invoked
inside a real application.

The operator layer begins with a call, not a lookup. The \aagent{} writes the
hot-operator case and invokes the \kagent{}. The map is the \kagent{}'s
own knowledge base, so whether a cell matches is decided inside the
\kagent{}. The \aagent{} sees only the validated kernel that comes back.
A returned cell may still be adapted. When no cell answers the case, the
\kagent{} forges one. Four conditions send a case down that forging path.
The region may need a single fused operator rather than the separate ones
the library holds. The application may use a non-standard primitive no cell
covers. The matching cell may exist only at another shape or precision. Or a
transplanted kernel may fail on the application's own data. The non-standard
primitive is the most common of the four across our candidates. The call
runs one way: the \aagent{} invokes the \kagent{}, never the reverse. The
\aagent{} also never reads the map. The knowledge base stays inside the
agent that owns it (Figure~\ref{fig:arch}).

The order is deliberate, for two reasons. First, where a stencil is fused
with neighboring operators, the fusion must come first. The operator's
boundary is not fixed until the fusion sets it. Second, operator forging's
end-to-end effect is bounded by Amdahl's law through the operator's fraction
$f$. Doing framework work first freezes $f$. The operator speedup is then
measured against a stable fraction, and its end-to-end ceiling shows
cleanly.

The operator layer is not limited to stencils. The \kagent{}'s rules contain
no stencil-specific domain knowledge. That knowledge lives only in its
knowledge base. The loop is therefore a general optimizer, and it can forge
simple non-stencil operators too. This matters because of the Amdahl
ceiling. Once the stencil region is much faster, its complement $1-f$
becomes the bottleneck. A non-stencil operator can then become the new
hotspot, and the same \kagent{} takes it on. The experiments report these
cases.
\section{The \sysname{} Harness}
\label{sec:harness}

\subsection{Positioning}
\label{sec:harness:positioning}
An agent that produces a speedup can also overstate it. \forge{} therefore
checks every claim in a \harness{} built from two gates, one per level:
the operator gate and the application gate. Their verdicts never mix. Both
require a fair comparison, an independent
correctness oracle, and honest aggregation and provenance. Each requirement
is an executable check on a measurement surface the agent cannot edit, making
an overstated speedup a system-level error rather than a question of trust.
We call this arrangement \emph{codified preclusion}. Every way to fabricate
a number is paired with the requirement that precludes it. Each requirement
is enforced in code before the run, not audited after it. The enforcement
needs no separate auditor. The manifest, the driver, the timer, and the
oracle sit outside the agent's write scope, so a run that violates a
requirement cannot produce a number at all. ForgeTrain keeps a review agent
that audits a round for proxy execution and fabricated metrics. Here the
same frauds are precluded by construction. What remains are the distortions
that carry a real program, a real dataset, and real timing yet still
mislead. The ablation of \S\ref{sec:harness:ablation} measures exactly
those. We present the application requirements according to the fraud each
blocks, so the protocol itself is testable.

\subsection{The operator gate}
\label{sec:harness:operator}
Every \kagent{} round must clear the operator gate, which runs on a surface
the loop does not control. The \emph{correctness} gate uses an independent oracle and a
scale-invariant, per-precision relative error. The tolerance cannot be
loosened case by case. The \emph{zero-regression} gate covers every accepted
case with a persisted prior-best baseline that advances only on acceptance.
That baseline is the loop's own history, distinct from the external SOTA
used for final evaluation. Both gates read an idle-GPU canonical
re-measurement instead of noisy in-loop timing. The agent edits only kernel
source. The driver, timer, and oracle remain outside its reach. Concrete
thresholds are reported with the experiments.

\subsection{The application gate}
\label{sec:harness:application}
A reported application speedup is admitted only if it satisfies the
requirements of Table~\ref{tab:goldstandard}, each of which blocks a specific
way to fabricate a favorable number.\label{sec:harness:goldstandard} These
fall on the three fronts above. The comparison rows force two variants of
the \emph{same real program}. The variants run on the program's own inputs,
compare against the baseline it ships, differ by a single switch, and are
timed fairly. The correctness row hands the verdict to the program itself.
The last two rows keep the reported number honest.
\begin{table}[t]
  \centering
  \caption{The requirements a trustworthy speedup must satisfy and the
    fraud each precludes.}
  \label{tab:goldstandard}
  \small
  \begin{tabular}{@{}p{0.46\linewidth}p{0.46\linewidth}@{}}
    \toprule
    Requirement & Fraud precluded \\
    \midrule
    Real upstream program (no proxy)
      & Synthetic or proxy benchmark as end-to-end \\
    GPU-version-first baseline
      & CPU straw-man comparison \\
    Program's own dataset
      & Cherry-picked stencil-heavy input \\
    Single switch, byte-identical \texttt{orig}
      & Asymmetric variants; weakened baseline \\
    Program's own correctness check
      & Self-written oracle; ``wrong but faster'' \\
    Program's own timing, interleaved median
      & Favorable windows; setup/IO in baseline \\
    All-dataset geometric mean
      & Single-dataset cherry-pick \\
    Enforced \texttt{REAL} provenance
      & Proxy or cross-architecture pass-off \\
    \bottomrule
  \end{tabular}
\end{table}

Correctness is settled by the application through one of three manifest modes.
\texttt{builtin} consumes the application's validation verdict through a result
pattern with an explicit failure pattern. \texttt{same\_program\_output\_diff}
compares both variants' numerical outputs on the same input.
\texttt{cross\_variant\_stdout} compares their matched output. The oracle
therefore always comes from the program itself.

Reporting follows two rules.\label{sec:harness:provenance} Multi-dataset
applications report the all-dataset geometric mean, and \texttt{REAL}
requires a same-architecture GPU measurement. Proxies are disallowed, and
cross-architecture candidates remain outside the headline. All \numapps{}
validated applications are therefore \texttt{REAL}.

Each rule is enforced in code before execution. A schema-checked
manifest declares the baseline, datasets, switch, correctness mode, and
provenance. A manifest without a real-program block is rejected, so an
operator proxy cannot pass as end-to-end. Provenance is parsed, not
trusted. Both variants must build and run before timing starts. The
\aagent{} edits only the candidate and its switch. The driver, timer,
oracle, byte-identical \texttt{orig} branch, and manifest stay outside its
reach.
\section{Experiments}
\label{sec:experiments}

We validate three claims.
\textbf{(1) Forged operators beat per-case SOTA.} Against the strongest
public baseline for each case, the forged library wins 37 of 37 cases
(\S\ref{sec:exp:operator}).
\textbf{(2) Application wins come from forging, not reuse.} Across
\numapps{} real codes the end-to-end median is \medspeedup{}, every win is a
per-application solution rather than a generic operator swap, and the gain
anti-correlates with baseline quality (\S\ref{sec:taxonomy}).
\textbf{(3) The numbers can be trusted.} The harness corrected our own
results, and ablating any single requirement distorts a real measurement
(\S\ref{sec:exp:integrity}).
Throughout, the two bodies of evidence stay strictly separate. The operator
library is validated only by kernel microbenchmarks, the \numapps{}
applications only through the application gate. Neither endorses the
other.

\subsection{Setup}
\label{sec:setup}
Headline numbers are taken on a single NVIDIA A100. The operator library is
additionally measured, and selectively re-forged, on an H100 and a B200.
Public baselines were rebuilt on the H100 only, so SOTA comparisons span
A100 and H100, while the B200 carries transfer results. Operator cases are
compared against the strongest public system per case. These are
Halide~\cite{ragankelley2013halide} and Devito~\cite{luporini2020devito}
across all types, plus EBISU~\cite{zhang2023ebisu},
ConvStencil~\cite{li2024convstencil},
FlashFFTStencil~\cite{han2025flashfftstencil}, and
DRStencil~\cite{you2021drstencil} on the types each supports.
Temporal-blocking baselines are compared only against our TB operators.
Speedups are
reported split by precision, so fp16 traffic savings cannot inflate a
same-precision headline. The operator gate admits a case only below a
scale-invariant relative error of $10^{-4}$ (f32/f64) or $10^{-2}$ (fp16,
within $5\times$ of the numerical floor, E4). The case must also fall within
a $2\%$
zero-regression band. Both readings come from the seven-pass idle-GPU canonical
measurement (\texttt{robust\_canonical.json}).
Applications are compared against their own GPU implementations. We lead
with the median speedup, because the geometric mean is lifted by a few
large outliers. Multi-dataset applications report the all-dataset geometric
mean. All timings use the seven-pass interleaved canonical protocol.
The raw variance analysis is provided in the supplementary material.

\subsection{Operator-Level Forging}
\label{sec:exp:operator}

\subsubsection{Breadth of differentiation}
\label{sec:kernel:breadth}
The forged library is broad. It contains \numkernels{} distinct
\texttt{\_\_global\_\_} kernels spanning nine stencil types, multiple grid
shapes, and f16/f32/f64 precisions. Of those, the \numdispatchconfigs{}
benchmarked configurations trigger \numdispatchkernels{}. The full dispatch
view is in the supplementary material, and we keep the two counts distinct
throughout. The specialization is structural. Within an operator the code
path branches by shape. \texttt{box\_2} dispatches three structurally
distinct kernels across cubes and slabs. \texttt{star\_1} routes cubes and
anisotropic slabs to different fp16 kernels. Across operators the
mechanism differs along the compute, memory-layout, and
time-depth axes.
The \texttt{diamond} family shows the harness consolidating rather than
multiplying kernels. Across roughly sixteen generated candidates it
selected one fused kernel serving the 4- and 8-step
depths, plus a separate single-step kernel. Blind per-configuration
generation would have emitted one kernel per configuration. Each branch is
admitted only through the operator gate's double gate (E5;
\texttt{csrc/stencil\_kernel.cu}).

\subsubsection{Library vs.\ per-case SOTA upper bounds}
\label{sec:kernel:vs-sota}
\begin{figure*}[t]
  \centering
  \includegraphics[width=\linewidth]{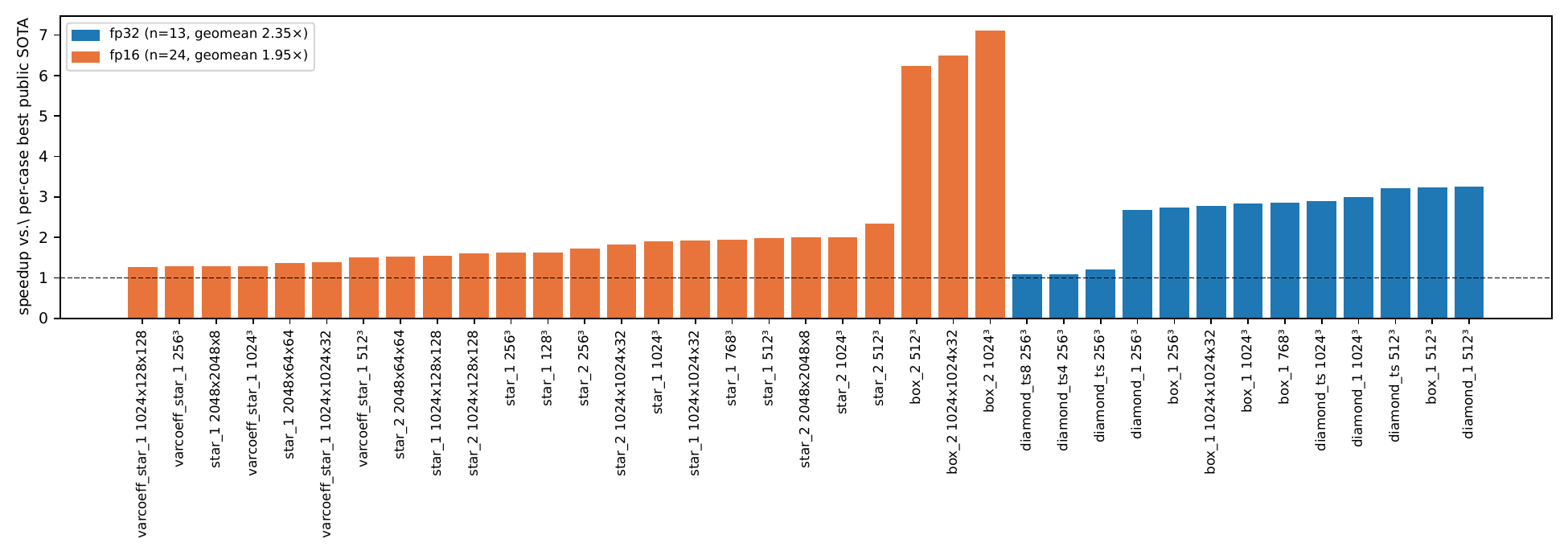}
  \caption{Library vs.\ per-case public-SOTA upper bounds, split by
    precision (blue f32, orange fp16; dashed line is parity). A100,
    microbenchmark scope (E3).}
  \label{fig:vs-sota}
\end{figure*}

Against same-precision f32-vs-f32 baselines ($n{=}13$), the library reaches
a geometric-mean speedup of \kernelgm{} (median \kernelmed{}). That is the
conservative headline, with a bootstrap $95\%$ confidence interval of
\kernelci{} over the seven-pass clean measurements. One
\texttt{diamond\_ts4} cell carries no baseline in the robust set, leaving
$n{=}12$. Against mixed-precision fp16-vs-f32 baselines
($n{=}24$) it reaches \kernelfphgm{} (median \kernelfphmed{}), reported
under the split-precision rule of \S\ref{sec:setup}. Across all 37 cases the combined geometric
mean is \kernelallgm{}, 37 wins and 0 losses (Figure~\ref{fig:vs-sota},
E3). The memory-bound operators run at \rooflinepeak{} of peak HBM
bandwidth (median \rooflinemedian{}). This also explains the narrower
variable-coefficient margin (\varcoeffgm{}, $1.27$--$1.51\times$, $n{=}5$).
Halide and Devito are themselves near that ceiling there. A narrow gain
therefore shows both sides near peak, not a weak operator. Every committed
fp16 kernel sits at the fp16 numerical floor, so the fp16 gains cost no
accuracy (E4). Tiny grids, where launch overhead dominates wall-clock, are
arbitrated on pure kernel time via \texttt{ncu} (for example \texttt{star\_1}
$128^3$: \arbwall{} wall-clock, \arbncu{} arbitrated;
full detail in the supplementary material).

\subsubsection{A refuted tensor-core mapping: forging a negative result}
\label{sec:kernel:tc-refutation}
The library's one compute-bound operator, \texttt{box\_2} (SM~$86\%$,
DRAM~$33\%$), is the case where a tensor-core (TC) mapping is most
attractive. Recent SOTA libraries cast such kernels onto matrix units via
\texttt{im2col}/Toeplitz packing (ConvStencil~\cite{li2024convstencil},
FlashFFTStencil~\cite{han2025flashfftstencil}). The forge loop tried that
path, and the harness rejected it. Table~\ref{tab:tc-refutation} traces the
attempts. A naive \texttt{im2col} runs $67\times$ slower than the winning
CUDA-core branch. An $N$-batched Toeplitz mapping closes that to $6.2\times$,
and a bank-conflict fix to $5.1\times$. Tensor-pipe utilization nevertheless
never
rises above \tcutil{}. The three variants converge on the same conclusion.
Engineering removes staging overhead without feeding the tensor cores, so
the ceiling is structural rather than an implementation defect.

\begin{table}[t]
  \centering
  \caption{The \texttt{box\_2} tensor-core attempts vs.\ the winning
    CUDA-core branch (A100 f32, $512^3$; \texttt{ncu}).}
  \label{tab:tc-refutation}
  \small
  \begin{tabular}{@{}lrrr@{}}
    \toprule
    Variant & Runtime & vs.\ warp16 & TC util \\
    \midrule
    \texttt{ko2i\_warp16} (CUDA core) & 1.251\,ms & 1.0$\times$   & --- \\
    naive \texttt{im2col} (TC)        & 84.3\,ms  & 67$\times$    & 2.1\% \\
    Toeplitz, $N$-batched (TC)        & 7.81\,ms  & 6.2$\times$   & 8.9\% \\
    \quad $+$ bank-conflict fix (TC)  & 6.42\,ms  & 5.1$\times$   & 10.8\% \\
    \bottomrule
  \end{tabular}
\end{table}

An analytical model confirms this. Dense packing of a sparse stencil
pays a density penalty $D\approx\boxdensitypenalty{}$, which puts the
shared-memory operand-feed floor above warp16's measured runtime, while raw
MAC throughput sits far below it. No bank-conflict or occupancy lever
can move that floor (full model in the supplementary material).

The claim is scoped. It is A100-specific, because the operand-feed floor
comes from this architecture's shared-memory bandwidth and MMA shapes. It is
also bounded. We do not claim that no TC formulation can ever win. It
sharpens the concurrent roofline-regime model of
Gu~\etal{}~\cite{gu2026tensorcores}. The full scoping argument is in the
supplementary material.

\subsubsection{Synthesis beats search: an equal-compute ablation}
\label{sec:exp:synth-vs-search}
A three-arm, equal-compute ablation isolates the three ways to reach a fast
kernel. The first arm is an autotuner sweeping launch configurations over a
fixed naive kernel. The second is a one-shot LLM generating once with no
measurement feedback. The third is the forge loop synthesizing iteratively
against the harness. The autotuner gains
little (\ablautolo{}--\ablautohi{}) and the one-shot LLM is
slower than naive (\abloneshotlo{}--\abloneshothi{}). The forge loop
reaches \ablforgelo{}--\ablforgehi{}, even though it is charged overheads the
other arms are not. The gain therefore comes from the
measure$\to$feedback$\to$re-synthesize loop, not from search or
one-shot generation alone. Full methodology and the per-stencil table are in
the supplementary material.

\subsubsection{The generality tax has an architectural axis: A100$\to$H100$\to$B200}
\label{sec:kernel:cross-arch}
The generality tax has a second axis. A kernel forged for one architecture
is itself a generic artifact relative to the next. The performance it
leaves on the table is the same tax along a different axis. Moved
unchanged to a real H100, the A100-forged production kernels transfer at a
\archbasegm{} geometric mean. That tracks the HBM3-over-HBM2 bandwidth
ratio. The specialization lead over Halide and EBISU holds and slightly
widens. Re-forging recovers additional headroom only where the
new hardware moves a kernel off its binding constraint, for instance an
occupancy retune on the one compute-bound operator. That is a config-level
change on H100. On a B200 it is a structural rewrite. Once the memory-bound
majority falls off the roofline, the same library transfers at \bgengm{}
across \bgencells{} cells. One rule covers both transitions. Re-forging
pays exactly where the bottleneck moved. The cost of capturing the gain
depends on whether the freed resource is a configuration knob or demands a
rewrite. The full round-by-round walkthrough, including the H100
\texttt{box\_2} decomposition and the B200 re-forge attempts, is in the
supplementary material.

The map also generalizes beyond benchmarked shapes. The cubic and
edge-shape geometric means are \gencubicgm{} and \genedgegm{}, respectively.
Per-operator details are provided in the supplementary material.

\subsection{Application-Level Forging: The Specialization--Generality Trade-off}
\label{sec:taxonomy}

Two findings make the specialization--generality trade-off concrete
across \numapps{} real applications. The observed
speedups arise from per-app specialized solutions rather than from a
uniformly faster generic operator. The realized gain also anti-correlates
with the generic baseline's quality.

\subsubsection{Speedup distribution}
\label{sec:taxonomy:distribution}
Across the \numapps{} validated applications the end-to-end speedup has a
median of \medspeedup{} and a geometric mean of \gmspeedup{}. Representative
cases, including the outliers that lift the mean, are
tabulated in the supplementary material. The distribution is broad:
11
applications below $1.05\times$, 46 in $1.05$--$1.5\times$, 22 in
$1.5$--$3\times$, 14 in $3$--$10\times$, 7 at $\ge10\times$
(Figure~\ref{fig:levers}). It also spans domains, with roughly $42\%$ drawn from
industrial codes. Gains from a single generic accelerator would cluster.
These span two orders of magnitude. The eleven
applications below $1.05\times$ cleared the harness's no-regression floor, which
requires $\ge0.98\times$, and the program's own correctness check. They
remain in the count.

\subsubsection{The wins come from forging}
\label{sec:taxonomy:levers}
\begin{figure}[t]
  \centering
  \includegraphics[width=0.95\linewidth]{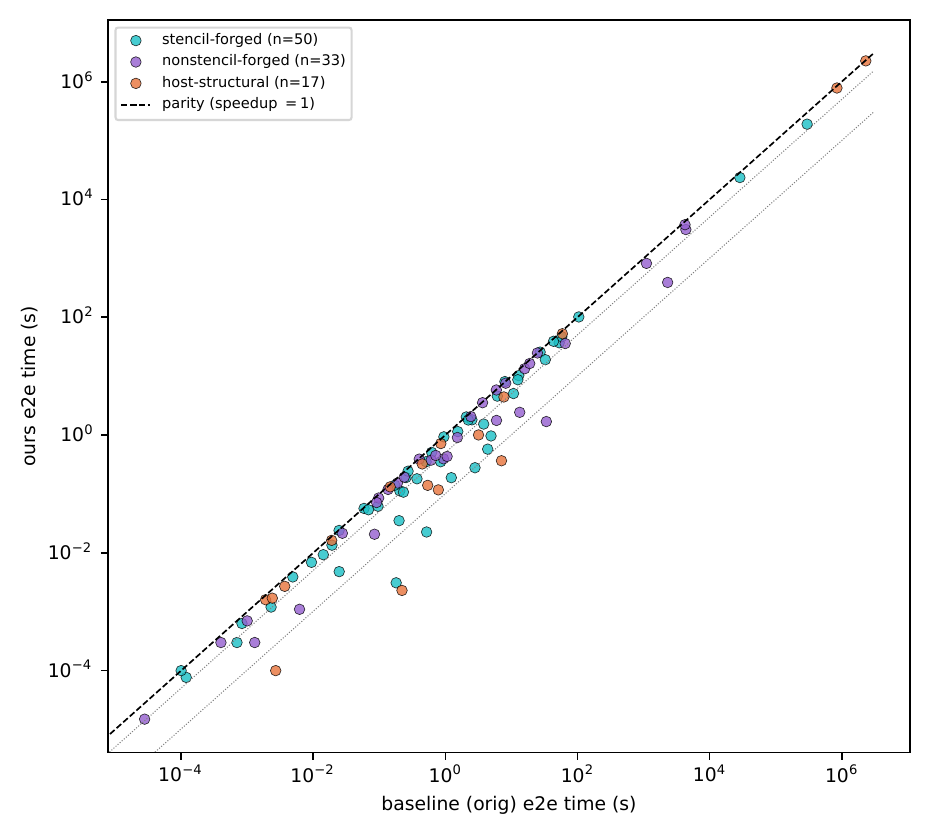}
  \caption{End-to-end runtime for all \numapps{} validated applications,
    colored by what was forged; below parity is a win (E2).}
  \label{fig:levers}
\end{figure}

Classifying the \numapps{} wins by how the speedup was produced provides the
most direct evidence for the paradigm (Figure~\ref{fig:levers}).
Fifty of the hundred came from a forged, bespoke stencil kernel: 32 written
for the application's shape and precision, plus 18 fusing the stencil with
the application's physics into a single kernel. Another 33 came from a forged
non-stencil
kernel, and 17 from a pure host or structural rewrite with no new device
kernel. None came from dropping in an off-the-shelf generic-library operator
unchanged. In practice, accelerating a real GPU application is rarely a
matter of calling a faster generic stencil.

We separate two kinds of win. Some of the largest factors repair
a naive upstream implementation, such as \texttt{haccmk} at $36.2\times$ and
\texttt{bspline\_vgh} at $58.39\times$. Their value is that the pipeline
finds and fixes them automatically, at scale, and under an enforced gate,
across all \numappstotal{} candidates. The case for the paradigm does not
rest on these outliers. The headline median \medspeedup{} is insensitive to
them by construction. On the 46 applications whose baseline was already
hand-tuned, the forged solution still wins on all 46 (median \genappmed{},
geometric mean \genappgm{}, minimum $1.002\times$). Each real code presents
shapes and domains the operators were never tuned for, and no
generic-library operator was reused. The end-to-end median is therefore
itself a measurement of generalization.

\subsubsection{The generality tax is measurable: speedup vs.\ baseline quality}
\label{sec:taxonomy:insights}
\begin{figure}[t]
  \centering
  \includegraphics[width=0.95\linewidth]{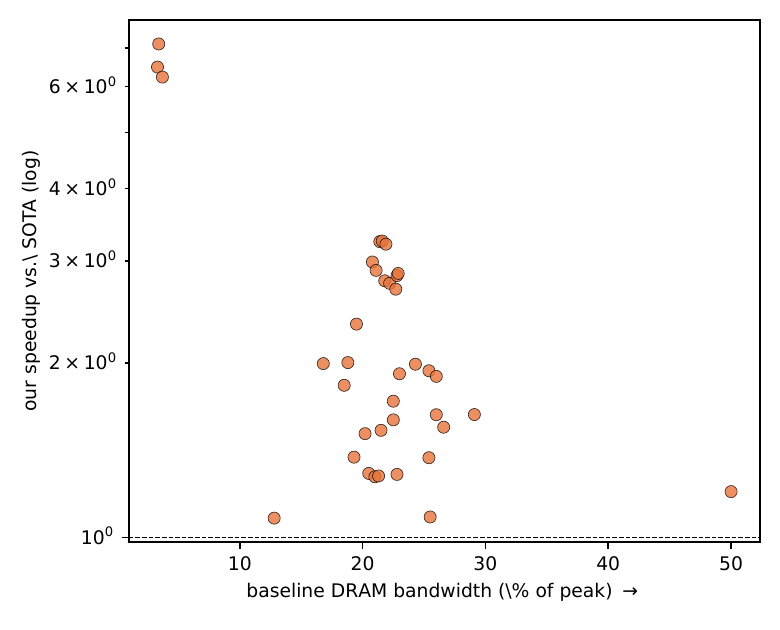}
  \caption{Speedup vs.\ baseline quality: operator-level Pearson $-0.60$
    ($n{=}37$) and application-level Spearman $-0.70$ ($n{=}43$).}
  \label{fig:baseline-quality}
\end{figure}

The second finding explains the distribution's shape. Realized speedup is
inversely correlated with the quality of the application's existing GPU
baseline (Figure~\ref{fig:baseline-quality}). It recovers a large factor
where upstream shipped a naive implementation, and lands near parity
where upstream was already well tuned. This makes the generality tax
measurable. The recoverable gain is set by that application's own baseline
rather than by any property of a shared operator. No single generic trick
could capture it.

\subsection{The Measurement Protocol in Action}
\label{sec:exp:integrity}

The harness is credible only if it changes numbers we would otherwise report.
We show it overturning our own results. The ablation then shows that
removing any single requirement distorts a real measurement.

\subsubsection{The protocol retracts single-dataset cherry-picks}
\label{sec:harness:retraction}
The protocol is enforced in practice. Requiring the all-dataset geometric
mean exposes both upward and downward single-dataset distortions
(Table~\ref{tab:retraction}). Each application's headline field now stores
that value.
\begin{table}[t]
  \centering
  \caption{Single-dataset results vs.\ the all-dataset values stored in the
    ledger.}
  \label{tab:retraction}
  \small
  \begin{tabular}{@{}lrr@{}}
    \toprule
    Case & Single-dataset & All-dataset geomean \\
    \midrule
    convolution3D & 2.08$\times$  & 1.358$\times$ \\
    sw4lite       & 1.003$\times$ & 0.998$\times$ \\
    cholla        & 1.294$\times$ & 1.021$\times$ \\
    laplace3d     & 1.628$\times$ & 2.241$\times$ \\
    minisweep     & 3.16$\times$  & 5.775$\times$ \\
    \bottomrule
  \end{tabular}
\end{table}
The same discipline rejected or downgraded \numrejected{} further
applications (supplementary material). The all-dataset rule binds wherever
an application ships more than one dataset, \numretracted{} of the
\numapps{}; the other 95 ship exactly one upstream dataset, so the fraud
this rule precludes, picking a favorable dataset, cannot arise for them.

\subsubsection{Measurement-protocol ablation}
\label{sec:harness:ablation}
Every ablated requirement distorts a real measurement, showing that none is
redundant (Table~\ref{tab:protocol-ablation}). The sharpest case is a
wrong-answer
\texttt{box\_2} reporting $13.2\times$ against an honest $6.23\times$.
\begin{table}[t]
  \centering
  \caption{Measurement-protocol ablation on real cases (E7).}
  \label{tab:protocol-ablation}
  \small
  \begin{tabular}{@{}p{0.34\linewidth}p{0.56\linewidth}@{}}
    \toprule
    Mechanism removed & Distortion on real cases \\
    \midrule
    All-dataset geometric mean
      & picks drift both ways, up to $+53\%$; one parity case becomes a fake
        win (Table~\ref{tab:retraction}). \\
    Real upstream program
      & synthetic loop: $2.03\times$; true e2e: $1.04\times$ (vpic FDTD,
        $f{=}6.8\%$), $1.96\times$ inflation. \\
    Symmetric-variant timing (byte-identical switch)
      & star\_1 $128^3$: symmetric \arbncu{} vs asymmetric \arbwall{},
        $1.87\times$ distortion. \\
    Program's own correctness
      & box\_2 $512^3$: honest $6.23\times$ vs wrong-but-faster $13.2\times$
        (floor $27.6\times$). \\
    Robust interleaved timing
      & star\_1 $768^3$: robust $1.94\times$ vs contaminated $0.80\times$
        ($2.4\times$ swing). \\
    \bottomrule
  \end{tabular}
\end{table}
\section{Conclusion}
\label{sec:conclusion}
The speedups reported here come from specialized solutions forged for concrete
cases rather than from any single generic method. \sysname{} makes this
practical with code-synthesis agents. The \kagent{} produces a
configuration-indexed kernel matrix that beats per-case public-SOTA upper
bounds in all 37 measured cases. The \aagent{} extends specialization to \numapps{}+ real
applications, where every win is a bespoke kernel or structural rewrite rather
than a generic operator swap.

These results rest on an agent-independent harness that enforces correctness,
fair measurement, and honest reporting, and that retracts even our own
favorable numbers when they fail. The results also show that specialized
structure transfers while its binding resource remains fixed, and that
re-forging pays when a hardware change moves that limit. Testing this rule
beyond stencil and broadening multi-dataset coverage are the next steps.

\bibliographystyle{ACM-Reference-Format}
\bibliography{references}

\appendix
\section{Extended Operator-Level Results}
\label{sec:supp:operator}

\subsection{Pure-kernel arbitration}
\label{sec:supp:arbitration}
Wall-clock speedup can be distorted on very small grids, where fixed launch and
timing overheads dominate the kernel itself. For \texttt{star\_1} at $128^3$ the
wall-clock ratio is \arbwall{}, a tiny-grid timing artifact; arbitrating with
Nsight Compute (\texttt{ncu}) on pure kernel time instead yields a \arbncu{}
win. We report the \texttt{ncu}-arbitrated figure for such cases and flag them,
which is itself an operator-level instance of the main paper's measurement
discipline. This is not a self-serving choice of the favorable number: this
exact case reappears in the main paper's measurement-protocol ablation as the
\emph{symmetric-variant timing} row, where the same \arbwall{} is shown to be
the \emph{distortion} (the number an attacker gets by letting one variant
pipeline launches while the other pays a per-call sync), and the symmetric
\texttt{ncu} \arbncu{} is the honest measurement. Reporting \arbncu{} here
and calling \arbwall{} a fabrication there are two sides of the same
claim.

\subsection{Dispatch breadth}
\label{sec:supp:breadth}
\begin{figure}[h]
  \centering
  \includegraphics[width=\linewidth]{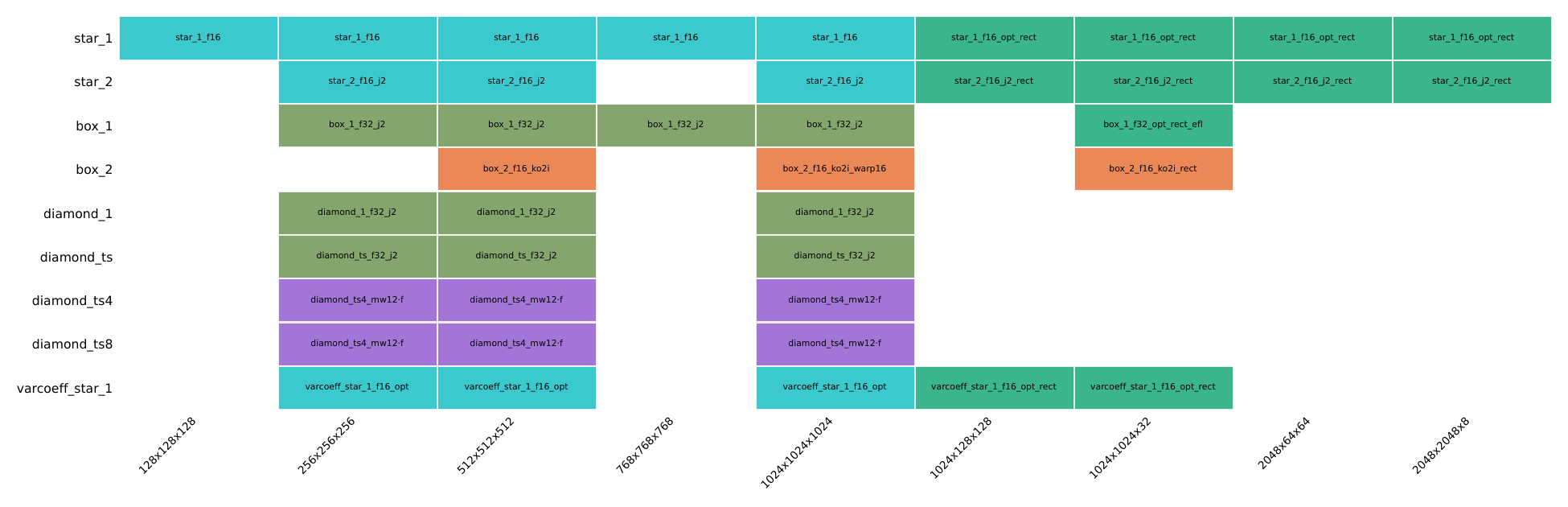}
  \caption{Specialization breadth as a dispatch view: each benchmarked
    configuration (type $\times$ shape $\times$ precision) is colored by the
    distinct kernel branch it dispatches to; no cell reuses a shelf-general
    kernel. A100.}
  \label{fig:breadth}
\end{figure}

Figure~\ref{fig:breadth} gives the dispatch view behind the main paper's
breadth counts.

\subsection{Per-case public-SOTA baseline identity}
\label{sec:supp:baseline-identity}
\begin{figure}[h]
  \centering
  \includegraphics[width=\linewidth]{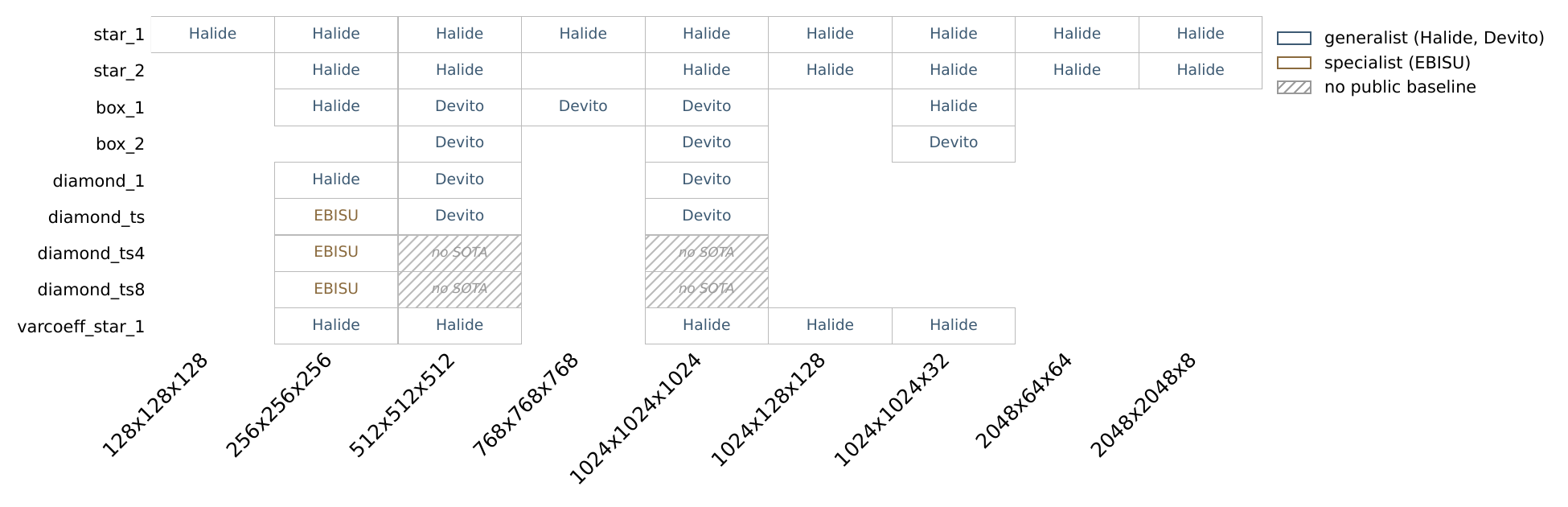}
  \caption{Identity of the per-case public-SOTA baseline, same (stencil,
    shape) axes as the main paper's dispatch-breadth figure: Halide and
    Devito (generalist, compared across all types) split the matrix, with
    neither dominating; EBISU (specialist) wins only the temporal-blocking
    cells it targets; hatched cells have no public baseline at all. A100.}
  \label{fig:supp-baseline-identity}
\end{figure}

The main paper's baseline methodology takes, for each case, the strongest
public system rather than one fixed competitor. Figure~\ref{fig:supp-baseline-identity}
gives the full per-cell mapping behind the main paper's precision-split
tally of which system that was.

\subsection{Tensor-core scope and relation to concurrent work}
\label{sec:supp:tc-scope}
Why the main paper's tensor-core ceiling is structural: a dense TC unit
doing a sparse computation pays a \emph{density penalty}
$D\approx\boxdensitypenalty{}$, the dense MACs it must issue per useful
output over the useful MACs (a naive $N$-waste packing is worse,
$D\approx16$). Two analytical floors scale with $D$ and order the wrong way
for the tensor cores: the raw MAC-throughput floor is only
$\approx0.71$\,ms, below the winning \texttt{warp16} branch, so compute is
\emph{not} the binding resource; the binding floor is operand feed,
streaming $A$/$B$ fragments out of shared memory, which lands at
$\approx1.4$--$2.8$\,ms once WMMA bank-conflict alignment is accounted for,
already above warp16's measured $1.251$\,ms. Bank-conflict and occupancy
levers can only close the gap to this floor, not pierce it: warp16
keeps the stencil's reuse in registers at near-unbounded bandwidth, while
any TC mapping must materialize that reuse through shared memory, and $D$
inflates that feed traffic into a wall.

The main paper's tensor-core refutation is bounded. The
\texttt{im2col}/Toeplitz route we forged is provably at its floor regardless
of further polish, and the SOTA TC library built for this packing regime,
ConvStencil, still loses to warp16 by \tcconvloss{}, but we do not claim no TC
formulation can ever win, since the theoretical minimum $D$ for this box is
$2$--$3$ against a $2\times$-win threshold of $D<2.86$, and we have not
refuted a packing at that edge. This sharpens rather than contradicts the
concurrent roofline-regime model of Gu~\etal{}, who find TC wins only where
deep temporal fusion or large radius pushes a kernel into a genuinely
compute-bound region~\cite{gu2026tensorcores}; our \texttt{box\_2} is
compute-bound for the opposite reason, an irreducible scalar reduction, so
packing it densely inflates $D$ into the same operand-feed wall. TC
amenability therefore depends on the \emph{source} of the compute bound, not
merely its presence, and the harness established that boundary by measurement
rather than by a prior model.

\subsection{Synthesis-vs-search ablation}
\label{sec:supp:synth-vs-search}
Forging concrete code is different in kind from searching a parameter space or
generating from a template. We test that claim directly with a three-arm,
equal-compute ablation that isolates the three ways to reach a fast kernel,
all measured under the same CUDA-event timer and correctness gate on A100
(f32; \texttt{notes/ablation\_synth\_vs\_search.py}). The arms are: an
\emph{autotuner} that sweeps launch configurations over a fixed naive kernel;
a \emph{one-shot LLM} that generates a kernel once with no measurement
feedback; and the \emph{forge} loop that synthesizes iteratively against the
harness (Table~\ref{tab:supp-synth-vs-search}).

\begin{table}[h]
  \centering
  \caption{Synthesis vs.\ search, equal-compute, f32 at $512^3$, speedup over the
    naive kernel. The forge arm is charged full dispatch$+$H2D overhead while the
    other arms are timed on kernel launch alone, so its lead is a conservative
    lower bound.}
  \label{tab:supp-synth-vs-search}
  \small
  \begin{tabular}{@{}lrrr@{}}
    \toprule
    Stencil ($512^3$) & Autotuner & One-shot LLM & Forge \\
    \midrule
    box\_1     & \ablautolo{}  & \abloneshothi{} & \textbf{\ablforgelo{}} \\
    diamond\_1 & \ablautohi{}  & \abloneshotlo{} & \textbf{\ablforgehi{}} \\
    \bottomrule
  \end{tabular}
\end{table}

Three findings follow. The autotuner reaches only \ablautolo{}--\ablautohi{}:
searching launch configurations over a fixed structure buys almost nothing.
The one-shot LLM is \emph{slower than naive}
(\abloneshotlo{}--\abloneshothi{}): generating once with no feedback does not
yield an optimized kernel, so the gain comes from the
measure$\to$feedback$\to$re-synthesize loop. Forge reaches
\ablforgelo{}--\ablforgehi{}, and this understates it, since the forge arm
pays full dispatch and host-to-device overhead while the other arms are
timed on kernel launch alone (steady-state pure-kernel figures are higher,
\eg $2.83\times$ for \texttt{box\_1} at $512^3$), consistent with the main
paper's independent SOTA comparison.

\subsection{Cross-architecture transfer (A100$\to$H100$\to$B200)}
\label{sec:supp:cross-arch}
We measure the architectural axis of the generality tax directly by taking
the A100-forged production kernels (unchanged, correctness-gated) and running
them on a real H100 (\texttt{sm\_90}, via a \texttt{cctl} devspace). Running
the loop on another generation is itself a configuration change: the GPU
model, compute capability, peak bandwidth, correctness oracle, and timer come
from a backend description rather than from the loop's instructions. Whether
to re-forge a given kernel there is a measured decision. The loop re-forges
only where profiling shows bandwidth headroom, occupancy that can be raised
without spilling registers, and an SM not already saturated by
instruction-level parallelism, and it leaves a kernel already bound at its
roofline alone.

Across 26 cubic cells the A100-forged library carries to H100 at a
\archbasegm{} geometric-mean speedup (\archbaselo{}--\archbasehi{}), tracking
the HBM3-over-HBM2 bandwidth ratio: the memory-bound kernels already saturate
the faster bus, so the forged structure transfers near-optimally. Re-running
the three-arm harness on H100 exposes where it does \emph{not} transfer for
free. The forge arm's speedup over naive \emph{shrinks}: \texttt{box\_1}
\archforgeboxa{}$\to$\archforgeboxh{}, \texttt{diamond\_1}
\archforgediaa{}$\to$\archforgediah{}, because the A100-forged structure is
no longer the optimal one on H100: it leaves recoverable headroom, which is
exactly the architectural generality tax made measurable. The autotuner, by
contrast, stays pinned near parity on H100 as it was on A100, and its best
launch block \emph{drifts} (\texttt{star\_1} $128\to160$): searching a fixed
kernel's configuration space hits its low ceiling on either architecture, so
the tax is not something a parameter search recovers. Recovering it is a
re-forge, which is cheap here (a routing/occupancy config change) and, on
hardware that relaxes a different bottleneck, can require a structural
rewrite.

What such a re-forge actually buys has to be checked in both directions: the
\texttt{box\_2} re-forge on H100 bundles two changes, re-selecting a
warp-level kernel and retuning occupancy, that read as a $7\%$ win combined
but come apart when measured on both architectures, since the warp-level
kernel is also faster on A100 ($1.09\times$ there, $1.03\times$ on H100, an
option the A100 campaign had missed rather than an H100 specialization) and
only the occupancy retune is architecture-specific ($+\bgenocch{}$ on H100
against $+\bgenocca{}$ on A100). We report the decomposition instead of the
combined number because the gate that produced it is the one the rest of the
paper rests on.

Crucially, the specialization advantage does not evaporate across the
generation: against Halide the forged kernels' edge holds and slightly widens
(geometric mean \archhalidea{}$\to$\archhalideh{} overall), and against
EBISU, the strongest temporal-blocking SOTA, our eight-step fused
\texttt{diamond\_ts8} still wins \archebisulo{}--\archebisuhi{} at
$256^3$/$512^3$ and completes correctly at $1024^3$ where EBISU fails to
launch. Taken over every H100 cell that carries a per-case public baseline,
the forged kernels hold a \archsotagmh{} geometric mean across
\archsotacells{} cells with no losses.

The next generation inverts the picture. Carrying the same forged kernels
from H100 to a B200 gives a geometric mean of \bgengm{} across \bgencells{}
cells, all clearing the correctness gate, against an HBM ratio of \bgenbw{};
the kernels are no longer riding the bus, since structures tuned to saturate
HBM3 under-fill HBM3e and the binding constraint moves from bandwidth to
latency and memory-level parallelism. Re-forge room reopens, and here the
capture is structural rather than a configuration knob: over \bgenrounds{}
gated rounds the loop produced one win, a deeper L2-prefetch sliding window
for \texttt{box\_1} worth $+\bgenforgehi{}$ at $512^3$, alongside several
recorded failures (\eg two \texttt{cp.async} formulations losing to barrier
overhead). Every B200 change is architecture- and shape-gated, so the A100
and H100 paths stay byte-identical.

One rule covers both transitions: re-forging pays where the new hardware
moves a kernel off its previous binding constraint, and the cost of
capturing the gain tracks whether the freed resource is exposed by a
configuration knob or demands a structural rewrite. H100 relaxed compute
while barely moving bandwidth, so the memory-bound majority transferred at
the bandwidth ratio and only the one compute-bound operator had anything to
recover, at config level; B200 widened bandwidth past what these kernels can
issue against, so the memory-bound majority fell off the roofline and
recovery meant restructuring how loads are kept in flight. One scope limit
applies: the public baselines were rebuilt on H100 but not on B200, so the
B200 numbers are transfer and re-forge measurements rather than a claim
against SOTA on that generation.

\subsection{How the specialization emerges}
\label{sec:supp:autonomy}
Specialization emerges incrementally: tracing the
loop's trajectory across its rounds (the canonical run spans R1--R321) shows
early rounds establishing a working kernel and later rounds forking it into
shape-specific branches as the harness rewards each divergence, including
disruptive refactors that run slower for several rounds before paying off
(Rule~14), a move a greedy, single-step search would never make. This is a
lightweight characterization of autonomy, not a causal attribution study
(\texttt{perf\_log.md}, E8).

\subsection{Generalization within the library: unseen shapes}
\label{sec:supp:generalization-op}
A natural objection is that the forged operators might be overfit to exactly
the shapes they were benchmarked on. They are not. Holding the operator
identity fixed, we compare each operator's speedup on the well-exercised
cubic shapes against its speedup on \emph{edge} shapes: long slabs and thin
rectangular domains that the tuning did not target. Aggregated across
operators, the cubic geometric mean is \gencubicgm{} and the edge geometric
mean is \genedgegm{}, a ratio of \genshaperatio{}: the unseen shapes do not
collapse. Even the weakest case, \texttt{box\_1} on thin slabs, still beats
SOTA at $2.70\times$ (versus $3.42\times$ on cubes). Because a single
operator's shape branches are selected by the harness and not hand-fit to a
benchmark, the specialization generalizes across shape rather than
memorizing the measured points.
\section{Applications the Framework Rejected or Downgraded}
\label{sec:appendix:rejected}
Alongside the \numapps{} validated integrations, the framework assigned a
non-success state to \numrejected{} further candidates rather than force each into
an \texttt{INTEGRATED} number: \texttt{RESEARCH}~6 (promising leads not yet
integrable), \texttt{OUT\_OF\_SCOPE}~5 (\eg PolyBench dense linear algebra,
retracted as outside the stencil premise), \texttt{DROPPED}~3 (abandoned after evaluation), and
\texttt{BLOCKED}~2 (\eg a refused Devito stand-in). The ledger thus records
\numapps{} integrated plus \numrejected{} non-validated candidates
(\numappstotal{} total). We include
this as backing detail for the main paper's retraction discussion, not as a
staged self-trial.

\section{Representative Benchmark Cases}
\label{sec:appendix:cases}
Table~\ref{tab:benchmarks} lists representative applications with the lever
behind each win. The large factors are launch-fusion and occupancy wins (\eg
\texttt{bspline\_vgh}/QMCPACK at $58.39\times$, \texttt{haccmk}/HACC at
$36.2\times$), not operator substitutions; pure stencil replacement sits at
the modest end (\texttt{fdtd\_em}/gprMax at $2.474\times$). Values are the
registry's all-dataset geometric means where an application ships multiple
datasets; only minisweep in this table does.
\begin{table}[htbp]
  \centering
  \caption{Benchmark applications with \texttt{win\_source}. Large
    speedups come from naive baselines and structural or launch-level rewrites,
    not from a faster generic stencil.}
  \label{tab:benchmarks}
  \footnotesize
  \begin{tabular}{@{}llr@{}}
    \toprule
    Application (registry name) & \texttt{win\_source} & Speedup \\
    \midrule
    QMCPACK (\texttt{bspline\_vgh})   & launch-fusion                 & 58.39$\times$ \\
    HACC (\texttt{haccmk})            & occupancy fix                 & 36.2$\times$ \\
    minisweep (\texttt{minisweep})    & structural                    & 5.775$\times$ \\
    hypre (\texttt{hypre})            & memory coalescing             & 3.857$\times$ \\
    gprMax (\texttt{fdtd\_em})        & true stencil replacement      & 2.474$\times$ \\
    FHd (\texttt{fhd})                & structural                    & 2.452$\times$ \\
    QuantLib (\texttt{bonds})         & structural (host-wrapper)     & 1.82$\times$ \\
    RTM (\texttt{rtm\_iso})           & stencil injection             & 1.808$\times$ \\
    HPCG (\texttt{hpcg})              & matrix-free SYMGS (bandwidth) & 1.712$\times$ \\
    \bottomrule
  \end{tabular}
\end{table}

\section{Experiment Manifest}
\label{sec:appendix:experiments}
Table~\ref{tab:experiments} maps each numbered experiment to the claim it
supports and to the artifact it is measured from. The E-numbers are the ones
cited throughout the paper.

\begin{table}[htbp]
  \centering
  \caption{Experiment manifest: claim, support, and source.}
  \label{tab:experiments}
  \footnotesize
  \begin{tabular}{@{}lp{0.26\linewidth}p{0.42\linewidth}@{}}
    \toprule
    \# & Supports & Source / status \\
    \midrule
    E1 & app taxonomy / headline & registry (median \medspeedup{} / geomean \gmspeedup{}) \\
    E2 & app taxonomy       & registry \texttt{win\_source}/\texttt{decision} \\
    E3 & operator library   & \texttt{robust\_canonical.json} \\
    E4 & gates + operators & \texttt{revalidate\_fp16\_full.py} (HEAD all PASS) \\
    E5 & operator library (main defense) & \texttt{custom\_stencil.py} /
                                  \texttt{stencil\_kernel.cu} (\emph{to compile}) \\
    E6 & gates + taxonomy & 5 retraction cases (registry) \\
    E7 & gates              & measurement-protocol ablation (5 rows, measured) \\
    E8 & operator library (main paper's autonomy discussion) & \texttt{perf\_log.md} (optional) \\
    \bottomrule
  \end{tabular}
\end{table}

\section{Evidence Map}
\label{sec:appendix:validation-map}
Table~\ref{tab:validation-map} states which gate validates which result.
Kernel microbenchmarks and the application gate never endorse each other's
numbers.
\begin{table}[htbp]
  \centering
  \caption{Which result is validated by which gate. The strict rule: no
    cross-endorsement between kernel microbenchmarks and the application
    gate.}
  \label{tab:validation-map}
  \small
  \begin{tabular}{@{}p{0.30\linewidth}p{0.26\linewidth}p{0.30\linewidth}@{}}
    \toprule
    Result & Validated by & Scope \\
    \midrule
    Operator-level forging   & Kernel microbenchmark  & A100/H100, per-case SOTA \\
    Application-level forging & Application gate      & \numapps{} real apps \\
    Speedup-source taxonomy  & Application gate      & \numapps{} real apps \\
    Measurement integrity    & Protocol (both levels) & enforced at both gates \\
    \bottomrule
  \end{tabular}
\end{table}

\clearpage

\end{document}